\documentclass[fleqn,10pt]{wlscirep}
\pdfoutput=1
\usepackage[utf8]{inputenc}
\usepackage[T1]{fontenc}
\usepackage{comment}
\usepackage[title]{appendix}
\usepackage{lscape}
\usepackage{geometry}
\usepackage{tabularx}
\makeatletter
\newcommand\setcurrentname[1]{\def\@currentlabelname{#1}}
\makeatother

\usepackage{xr}
\makeatletter
\newcommand*{\addFileDependency}[1]{
  \typeout{(#1)}
  \@addtofilelist{#1}
  \IfFileExists{#1}{}{\typeout{No file #1.}}
}
\makeatother

\newcommand*{\myexternaldocument}[1]{
    \externaldocument{#1}
    \addFileDependency{#1.tex}
    \addFileDependency{#1.aux}
}

\myexternaldocument{supplementary_material}

\title{A study on group fairness in healthcare outcomes for nursing home residents during the COVID-19 pandemic in the Basque Country}

\author[1,*]{Hristo Inouzhe}
\author[2,1]{Irantzu Barrio}
\author[1,3]{Paula Gordaliza}
\author[4,5]{María Xosé Rodríguez-Álvarez}
\author[6]{Itxaso Bengoechea}
\author[7,8,9]{José María Quintana}
\affil[1]{BCAM - Basque Centre for Applied Mathematics, Bilbao, 48009, Spain}
\affil[2]{Universidad del País Vasco (UPV/EHU), Department of Mathematics, Leioa, 48940, Spain}
\affil[3]{Universidad Pública de Navarra, Department of Statistics, Informatics and Mathematics, Pamplona, 31006, Spain}
\affil[4]{CINBIO, Universidade de Vigo, Department of Statistics and Operations Research, Vigo, 36310, Spain}
\affil[5]{CITMAga, Galician Center for Mathematical Research and Technology, Santiago de Compostela, Spain}
\affil[6]{Hospital Galdakao-Usansolo, Hospital at Home Unit, Galdakao, 48960, Spain}
\affil[7]{Hospital Galdakao-Usansolo, Unidad de Investigación,
Galdakao, 48960, Spain}
\affil[8]{Red de Investigación en Servicios Sanitarios y Enfermedades
Crónicas (REDISSEC), Galdakao, Spain}
\affil[9]{Red de Investigación en Cronicidad, Atención Primaria y
Promoción de la Salud (RICAPPS), Bizkaia, Spain}

\affil[*]{hinouzhe@bcamath.org}

\begin{abstract}
We explore the effect of nursing home status on healthcare outcomes such as hospitalisation, mortality and in-hospital mortality during the COVID-19 pandemic. Some claim that in specific Autonomous Communities (geopolitical divisions) in Spain, elderly people in nursing homes had restrictions on access to hospitals and treatments, which raised a public outcry about the fairness of such measures. In this work, the case of the Basque Country is studied  under a rigorous statistical approach and a physician's perspective. As fairness/unfairness is hard to model mathematically and has strong real-world implications, this work concentrates on the following simplification: establishing if the nursing home status had a direct effect on healthcare outcomes once accounted for other meaningful
patients' information such as age, health status and period of the pandemic, among others. The methods followed here are a combination of established techniques as well as new proposals from the fields of causality and fair learning. The current analysis suggests that as a group, people in nursing homes were significantly less likely to be hospitalised, and  considerably more likely to die, even in hospitals, compared to their non-residents counterparts during most of the pandemic. Further data collection and analysis are needed to guarantee that this is solely/mainly due to nursing home status.
\end{abstract}
\begin{document}

\flushbottom
\maketitle
%
%
%
%
\section*{Introduction}

The COVID-19 pandemic has put to the test the capabilities of our socio-economic systems. Chain supplies, labour markets, health care systems, and many others have suffered severe stress due to extremely atypical conditions. The extraordinary circumstances forced decision-making that was not standard and for which there often were no firmly established protocols. For instance, public healthcare systems were faced with huge burdens while material and personnel were scarce. This led to situations where life-or-death choices were forced upon medical staff. In this unprecedented environment, some vulnerable social groups were especially exposed. Amnesty International Spain claims that people residing in nursing homes in Spain (`residencias de ancianos' in Spanish) were unprotected and discriminated against during the COVID-19 pandemic\cite{AISpain:2020}. In hindsight, and with the worst of the pandemic behind, we want to apply rigorous statistical methods to evaluate if some of the assertions of Amnesty International Spain are also applicable to
the Basque Country, one of the Autonomous Communities (geopolitical divisions) in Spain.

The main objective of this work is to study whether there have been statistically meaningful differences between nursing home residents and non-residents, seen as demographic groups, in health care access, i.e., hospitalisation, and in health care outcomes, such as mortality and in-hospital mortality. 
These sorts of questions are particularly relevant nowadays in the fair learning research in Artificial Intelligence, where the word bias is used to refer to an inclination or prejudice for or against one person or group based on their characteristics. When such subgroups are related to sensitive information about individuals, for example being a nursing home resident (other examples of sensitive variables could be gender, race, age, etc. ), the concept of \textit{fairness} as the absence of such bias is raised. However, to avoid ambiguity in the rest of the paper, we will rather use the term 
(un)fairness, leaving the term bias for the concept of statistical bias. This statistical bias can come either from the bias of the estimator of the fairness measure or from the bias generated in the sample by unobserved variables.

With the widespread use of predictive algorithms, the mathematical formalisation of fairness has been one of the main challenges in scientific research recently. It is one of the fundamental points of contention where ethical considerations come into play. In particular, Machine Learning (ML) community has devoted great efforts to studying the possible sources of discrimination with respect to sensitive characteristics in the outcomes of automatic decisions made by algorithms and to developing mechanisms to mitigate or reduce its prejudicial impact\cite{Fazelpour2021, oneto2020fairness, Gordaliza2020}.  In general, algorithmic fairness objectives can be categorised into individual and group fairness. On the one hand, \textit{individual fairness}\cite{dwork12fairness,kusner2017counterfactual,heidari2018fairness} try to achieve ``similar behaviour towards similar individuals''.
On the other hand, \textit{group or statistical fairness} focuses on reducing inequalities at a group level, where groups may be defined using sensitive variables such as race, gender, age, or disabilities. The equity objective is expressed through a measure of statistical independence between the variables involved in the learning process.\cite{kamiran2009classifying,kamishima2011fairness,gordaliza2019obtaining,hardt2016equality}

In the present observational study, the goal is to assess the existence of (statistical, algorithmic) unfairness in past decisions made by humans. In the following, we drop the terms statistical an algorithmic and just use (un)fairness whenever it is clear from the context. While the question of whether decision-making algorithms should be held to higher standards of transparency than human beings\cite{gunther2022algorithmic} has been raised, similar definitions of algorithmic fairness can be also applied in this context. In particular, a causal-based approach is especially convenient when the sensitive attributes are assumed to be highly correlated with the rest (or at least, part) of the covariates (variables) in the dataset. Precisely, we settle for a tractable and common sense notion of (statistical) fairness as follows. Let $Y$ represent the binary outcome of interest, that is, hospitalisation, death, or death in hospital. Take $Z$ to be the binary nursing home status, i.e., if a person is ($Z=1$) or is not ($Z=0$) a resident in a nursing home. Finally, let $\mathbf{X}$ be the vector of covariates that could affect (or be affected by) both the outcome of interest and the nursing home status. An example of such covariates are age, comorbidities, time of hospitalisation, etc. In this setting, one can define \emph{fairness} as
\begin{equation}\label{eq:fairness}
    P(Y|Z, \mathbf{X}) = P(Y|\mathbf{X}),
\end{equation}
i.e., as the conditional independence $Y\perp Z|\mathbf{X}$. Hence, a fair conduct corresponds to the situation where once the covariates $\mathbf{X}$ are determined and accounted for, the nursing home status $Z$ is irrelevant for the outcome $Y$ (e.g., hospitalisation). However, dependence between $Z$ and $Y$ mediated by $\mathbf{X}$ is considered legitimate and can be present. Consequently, our statistical fairness definition is a proxy of the ethical notion that sensitive attributes (gender, sex, nursing home status) should not have direct effects on outcomes/decisions. Under the usual assumptions of the field of causal inference\cite{pearl2009causality, rubin_2006, imbens2015causal}, unconfoundedness, positivity, consistency, and no interference, (\ref{eq:fairness}) is equivalent to a lack of causal effect from $Z$ to $Y$ in the presence of confounders $\mathbf{X}$. This can be stated in terms of Average Treatment Effect (ATE) as follows
\begin{equation}\label{eq:fairness_ate}
    \mathrm{ATE} = E_\mathbf{X}\big(E(Y|Z=1,\mathbf{X}) - E(Y|Z=0, \mathbf{X})\big)=0.
\end{equation}

Equipped with (\ref{eq:fairness_ate}) as a notion of fairness, one can use techniques such as inverse propensity score weighting \cite{lunceford2004stratification, austin2011introduction, garrido2014methods}, sample matching\cite{rubin_2006}, and some ML methods\cite{neal2020introduction} to estimate (\ref{eq:fairness_ate}) for the data at hand. These are the basis for the procedure introduced in this work which is as follows. Treat the data as coming from an observational study, compute an estimate of the ATE (\ref{eq:fairness_ate}) and a confidence interval (CI) for it, and proceed as follows:
\begin{itemize}
    \item if 0 is inside the CI, there is \emph{no evidence} in the data \emph{of unfairness} due to nursing home status,
    \item if  0 is not inside the CI, there is \emph{evidence} in the data \emph{of unfairness} due to nursing home status.
\end{itemize}

A difficulty is the robust estimation of the ATE (\ref{eq:fairness_ate}). As conclusions on fairness stem from this estimate, we need some evidence that using different methods to estimate (\ref{eq:fairness_ate}) leads to concordant results. Hence, we need to account for the possible (statistical) bias of the employed estimators. 
For that reason, we chose to use a battery of estimators for (\ref{eq:fairness_ate}), with their main differences summarised in Table \ref{table:characterization_estimations}. More details are provided in \nameref{sec_Methods}. Accordingly, if a clear majority of ATE estimates provide evidence of unfairness, one can say that there is robust evidence of unfair conduct.

A limitation of this approach comes from the hypothesis of unconfoundedness, which states that all possible confounders (covariates that affect both $Y$, the outcome, and $Z$, the nursing home status) are accounted for, i.e., are part of $\mathbf{X}$. However, there are ways of assessing the effect of potentially unobserved covariates on the outcome and the nursing home status required to produce enough bias in the estimate of the ATE to overturn our evidence on (un)fairness\cite{veitch2020sense}. This can be used by medical experts to determine the plausibility of such unobserved covariates, hence partially alleviating the limitation of unconfoundedness.
\begin{table}[h!]
\centering
\begin{footnotesize}
\begin{tabular}{cccccc}
                    & Matching & Weighted samples & Estimation of $E(Y|Z,\mathbf{X})$ & Estimation of $E(Z|\mathbf{X})$ & Distance         \\ \cline{2-6} 
Unmatched           & No       & No               & No                     & No                   &                  \\ \hline
Unmatched 2         & No       & No               & Yes                    & No                   &                  \\ \hline
Matched Euc         & Yes      & Yes              & No                     & No                   & Euclidean        \\ \hline
Matched Euc 2       & Yes      & Yes              & No                     & No                   & Euclidean        \\ \hline
Matched Prop        & Yes      & Yes              & No                     & Yes                   & Propensity Score \\ \hline
Matched Prop 2      & Yes      & Yes              & No                     & Yes                   & Propensity Score \\ \hline
Inverse Weighting & No       & Yes              & No                     & Yes                  &                  \\ \hline
Inverse Weighting 2 & No       & Yes              & Yes                    & Yes                  &                  \\ \hline
\end{tabular}
\end{footnotesize}
\caption{The Unmatched estimation corresponds to treating the data as a randomised study, where the ATE is just the difference in the probability of the outcome between residents and non-residents. Unmatched 2 estimates (\ref{eq:fairness_ate}) by using ML to estimate the conditional expectation $E(Y|Z, \mathbf{X})$. Matching methods aim to produce (weighted) subsamples of the original data such that residents and non-residents as groups are as similar as possible with respect to some distance measure on the confounders. Inverse weighting relies on estimates of the Propensity score, $E(Z|\mathbf{X})$, to produce weighted samples with a similar objective. Further details are given in \nameref{sec_Methods}.}
\label{table:characterization_estimations}
\end{table}

To summarise, in this work we study unfairness, understood as a causal effect of nursing home status on relevant healthcare outcomes in the presence of confounders, with a methodology that tries to account for bias in the estimation both due to the estimators and the possible unobserved variables. In this way, our results are more likely to reflect reality and not mere data artefacts. 


The rest of the paper is organised as follows. In section \nameref{sec_Results}, we present the main results of our inquiry providing evidence of unfairness towards nursing home residents and evaluating the quality of the ATE estimates. Afterward, we continue with a \nameref{sec_Discussion} on the main results. We end with a \nameref{sec_Methods} section where details on the data and its collection are provided, followed by a comprehensive overview of the methodology and the required techniques. There we present our novel method of estimating ATE based on matching samples by trimmed optimal transport and provide some novelties to the ATE diagnostics workflow.

\section*{Results}\setcurrentname{Results}\label{sec_Results}
\begin{figure}[h!]
    \centering
    \includegraphics[scale=0.29]{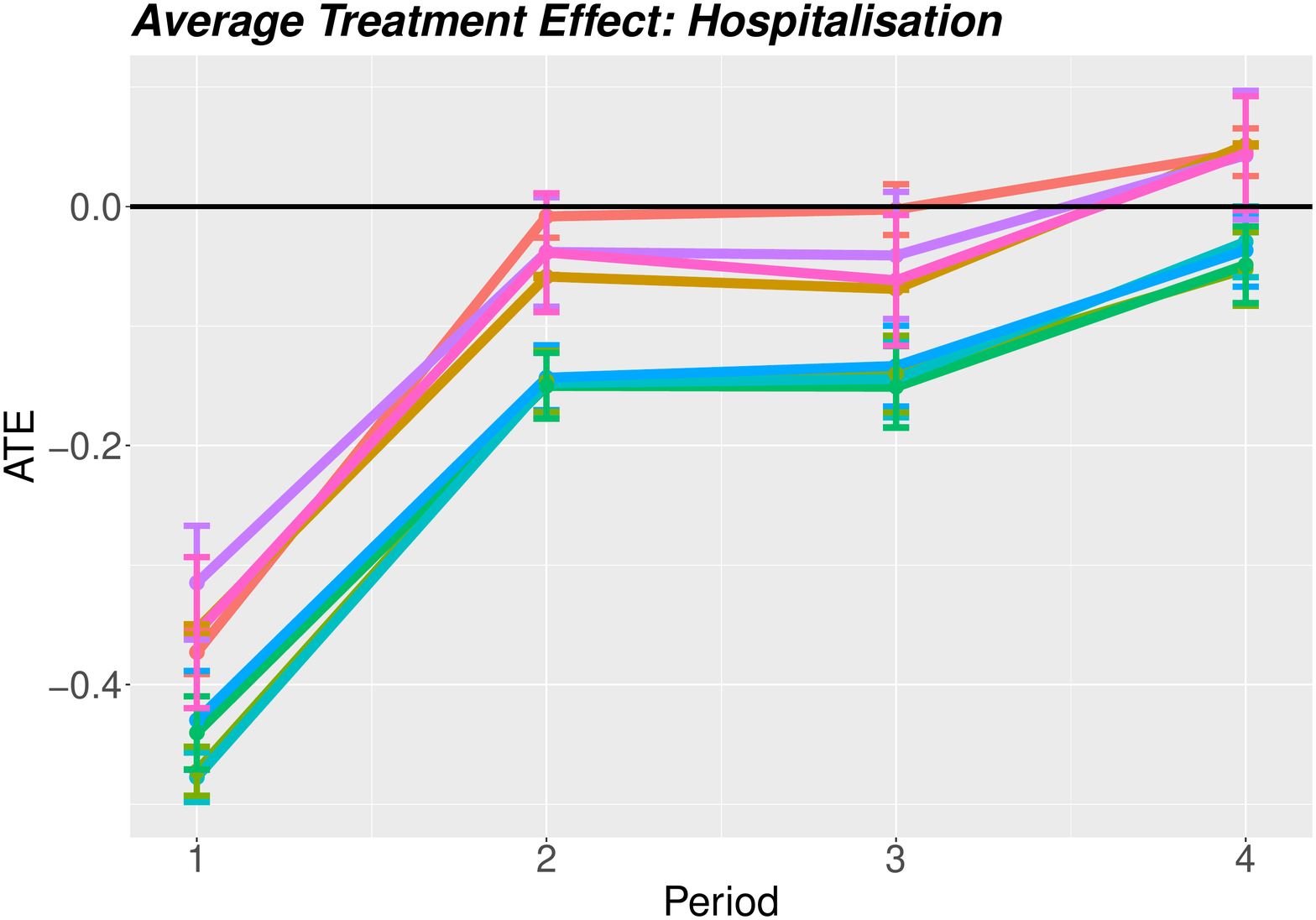}\includegraphics[scale=0.29]{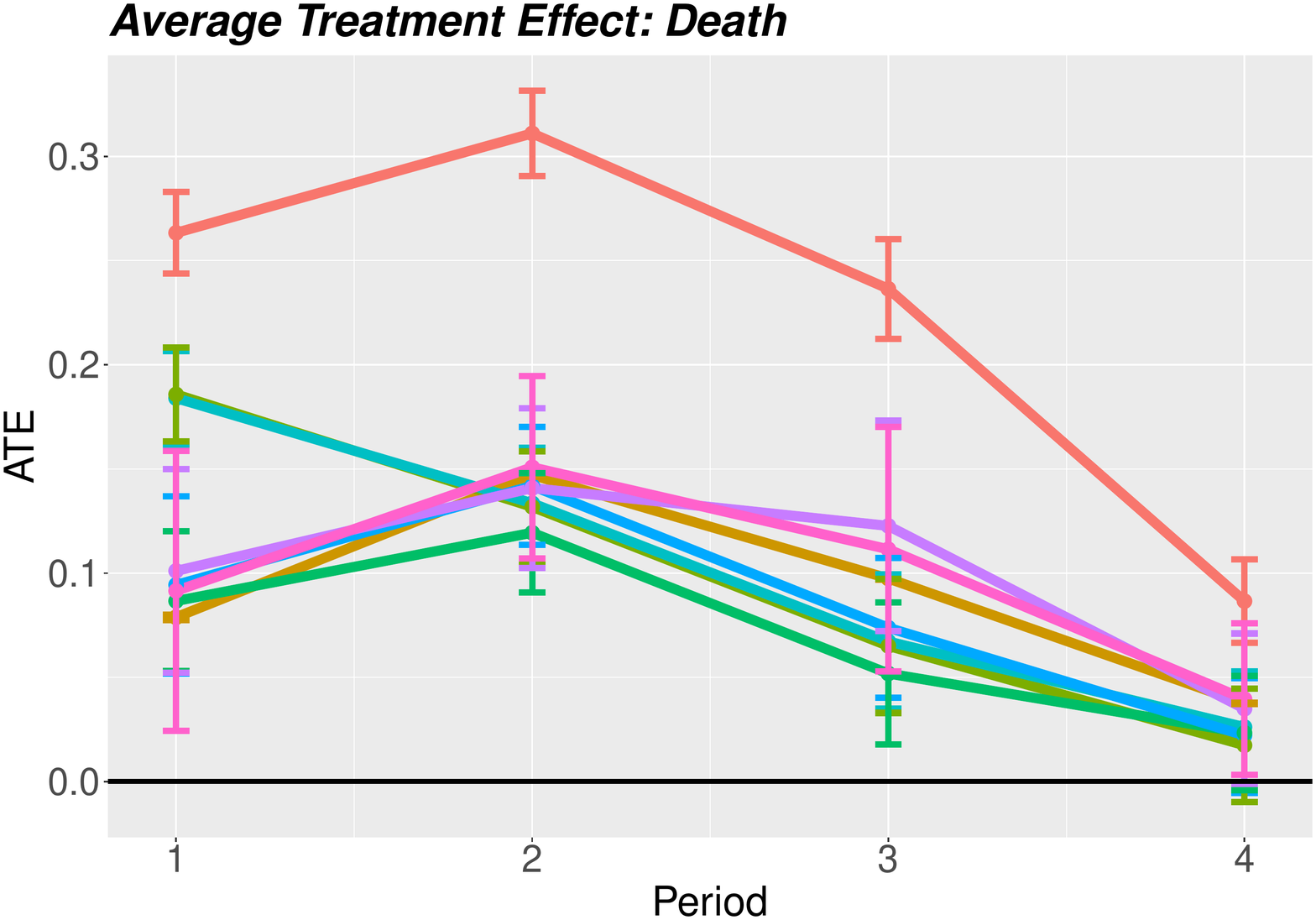}
    \includegraphics[scale=0.29]{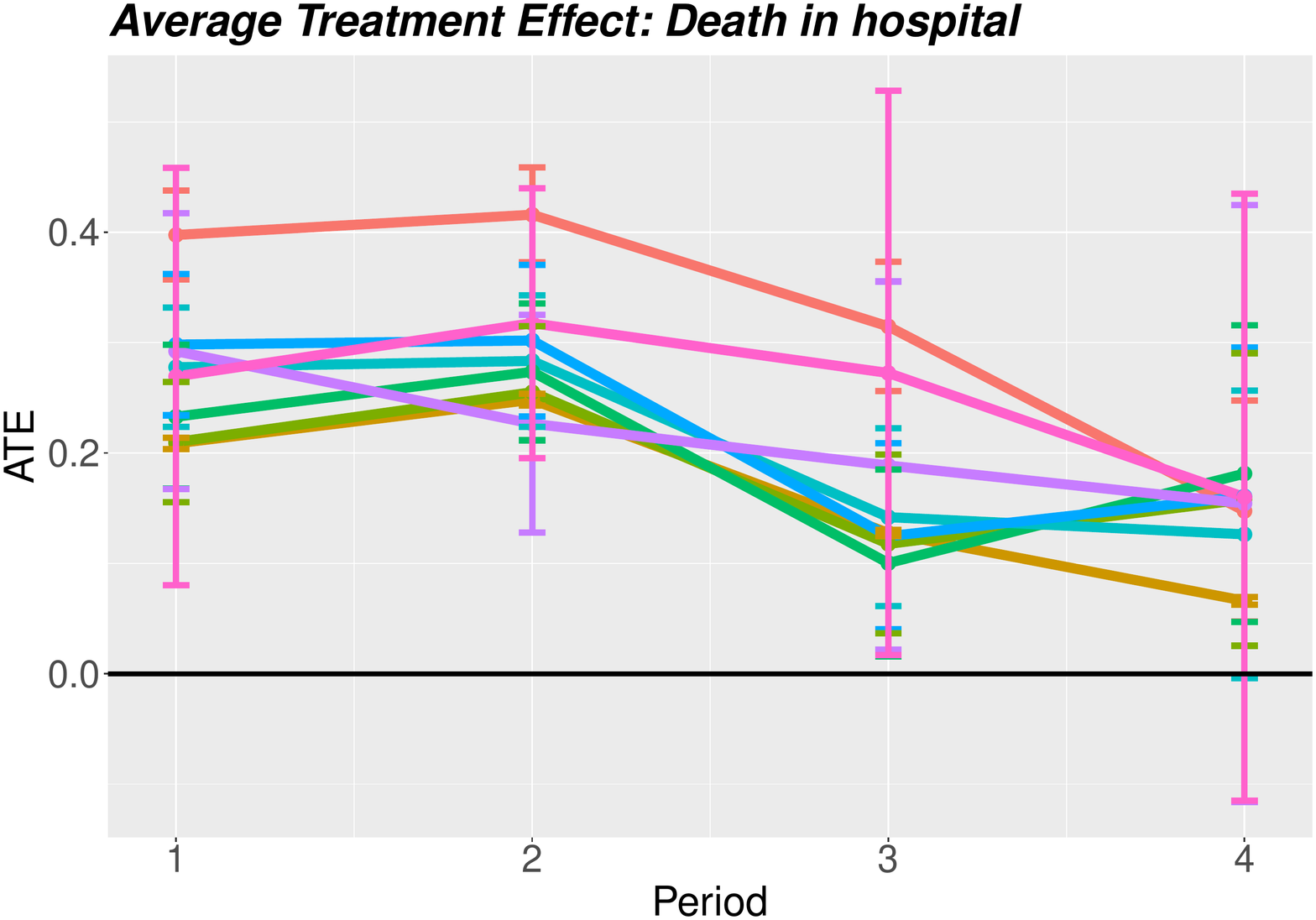}\hspace*{10pt}\includegraphics[scale=0.37]{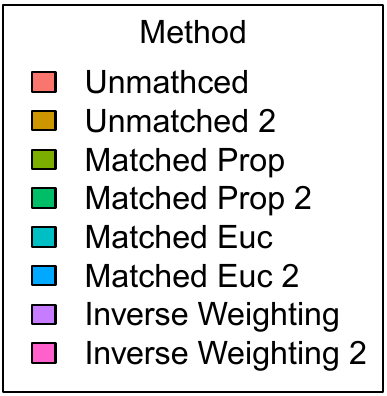}
    \caption{ATE estimation for four different periods, eight different estimators, and three different outcomes: hospitalisation, death, and death in hospital. Vertical error bars represent the 95\% confidence intervals.}
    \label{fig:ate_all_periods}
\end{figure}
In this section, we lay out the main results of our study concerning the ATE estimation in subsection \nameref{sec_Discrimination} and an evaluation of the ATE estimates in subsection \nameref{sec_Quality}. Firstly, a short overview of the dataset in use is supplied. Data consisted of individuals with a confirmed positive test for COVID-19 with sociodemographic information, baseline comorbidities and time information about hospitalisation and testing (when a SARS-COV-2 positive test was provided). Hospital admission and death due to COVID-19 were defined following Portuondo-Jiménez \textit{et al.} (2023) \cite{portuondo2023clinical}. Details on the data and the collection process are provided in the subsection \nameref{sec_Data} in \nameref{sec_Methods}. The main outcomes of interest, $Y$, were: hospitalisation, death, and death in hospital. The confounders we considered, a portion of all the availables, were: 
\begin{align}
\label{eq:covariates}
    \mathbf{X}=\{&\text{age, Charlson index, time until (SARS-COV-2 positive) test, time until hospitalisation, time from positive}\nonumber\\
    &\text{to hospitalisation, number of (baseline) prescribed treatments, sex, cancer, respiratory disease, cardiopathy,}\\&\text{ heart failure, respiratory illness, liver disease, dementia}\}.\nonumber
\end{align}
Only patients aged over 60 years were included in the analyses since the vast majority of residents of nursing homes in the Basque Country were part of that age group. 

The data were divided into four relevant periods, corresponding to different stages of the COVID-19 pandemic in the Basque Country:
\begin{center}
    \textbf{Period 1:} 1/3/2020 - 30/6/2020 \quad \textbf{Period 2:} 1/7/2020 - 31/12/2020\\
	\textbf{Period 3:} 1/1/2021 - 13/12/2021\quad \textbf{Period 4:} 14/12/2021 - 9/1/2022.
\end{center} 
Broadly speaking, Period 1 corresponds with the first wave of the pandemic. Period 2 comprises several waves previous to the start of vaccination and the dominance of the Delta variant, which belong to Period 3. Lastly, Period 4 is characterised by  the beginning and dominance of the Omicron variant. This periodisation equates to an effort to homogenise confounding in the data since smaller time periods corresponded to a more similar epidemiological situation than the one of the data as a whole.

\subsection*{Unfairness due to nursing home status}\setcurrentname{Unfairness due to nursing home status}\label{sec_Discrimination}

Figure \ref{fig:ate_all_periods} depicts, for each period and the different outcomes considered in this study, the ATE point estimates and 95\% bootstrap confidence intervals, using the estimators shown in Table \ref{table:characterization_estimations}. The plots should be interpreted as follows. If the error bar (i.e., 95\% confidence interval) for the corresponding period and estimator crosses 0 (marked in solid black), there is no evidence of unfairness. The tables used to generate Figure \ref{fig:ate_all_periods} are available in Appendix \ref{Ate_table}, where one can see detailed values which help to better interpret the plots, particularly near zero. Negative values of ATE in hospitalisation mean that nursing home residents were less likely than non-nursing home residents to be hospitalised. Positive values of ATE in death and death in hospital convey that nursing home residents were more likely to die, both in general and in hospitals, than non-nursing home residents.

With this in mind, one can see that for all outcomes, and periods 1-3, a clear majority (more than 60\% or all) of the estimators provide evidence of inequities, with negative consequences for the group of nursing home residents. While for the last time period, there is no such unanimous evidence of unfairness, for the first period, there is not only unanimity but also a considerable amount of unfairness. From Figure \ref{fig:ate_all_periods}, one can see that the size of unfairness changes with the time periods and the type of estimators. The latter is particularly notorious in hospitalisation, where there seems to be two groups of estimators, the ones obtained through matching showing a significant effect on periods 2 and 3, and the rest showing smaller or no effect on the same periods. The unfairness' size is particularly relevant because of the confidence bounds closeness to zero in period 2 for hospitalisation and in period 3 for all outcomes. These combinations (of periods and outcomes) are the most susceptible to be affected by bias due to unobserved variables. 
\begin{figure}[h!]
    \centering
    \includegraphics[scale=0.4]{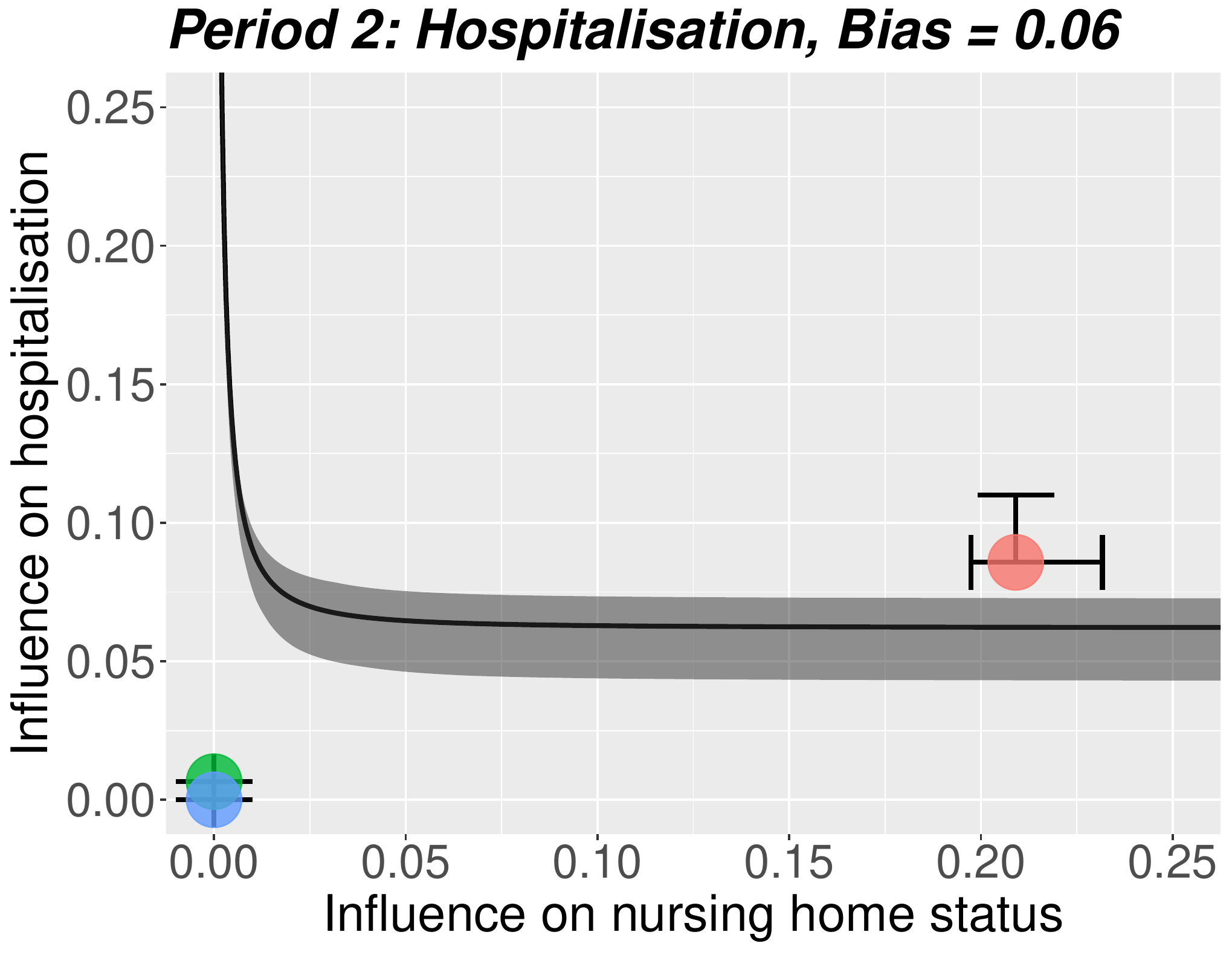}\includegraphics[scale=0.4]{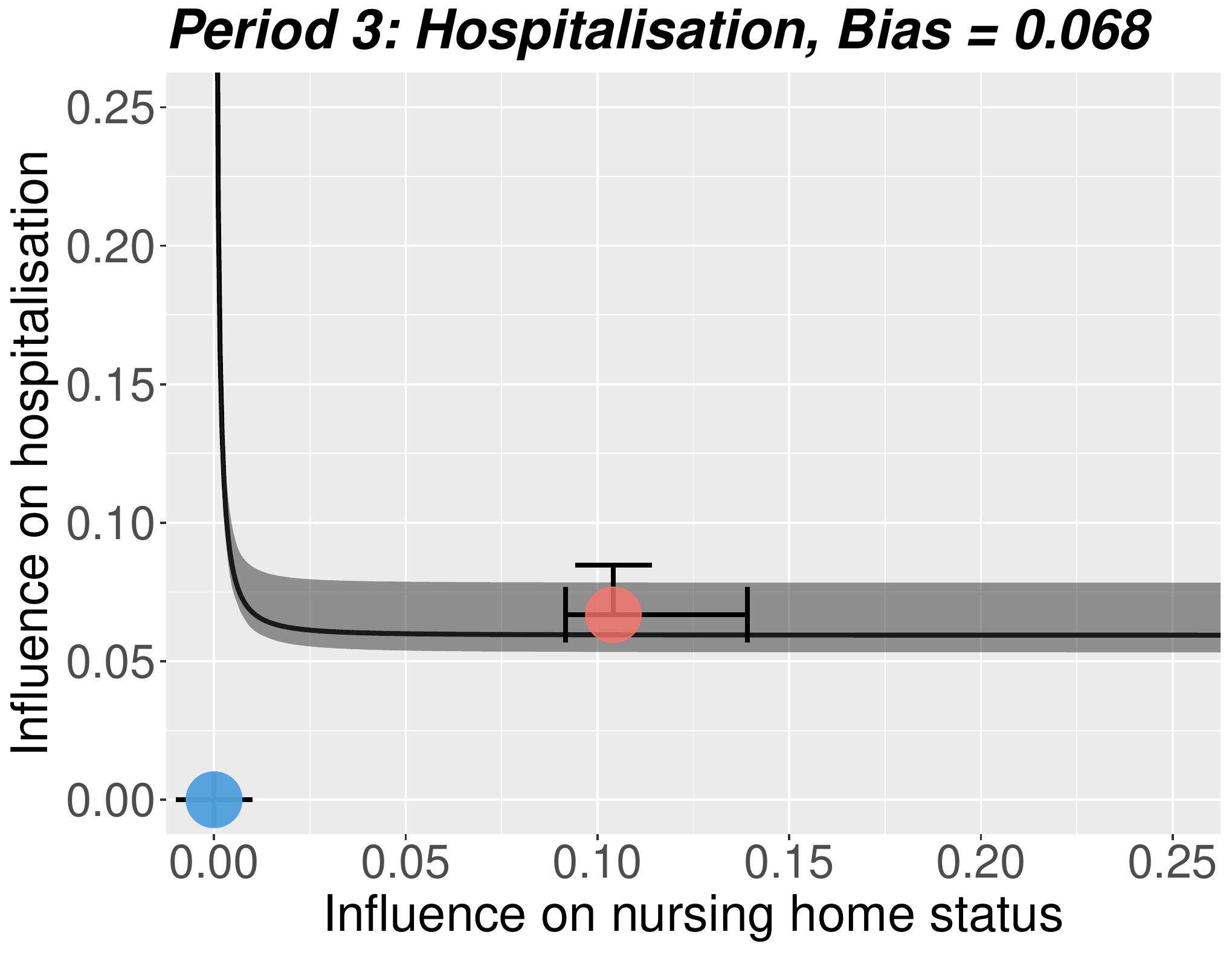}
    \includegraphics[scale=0.4]{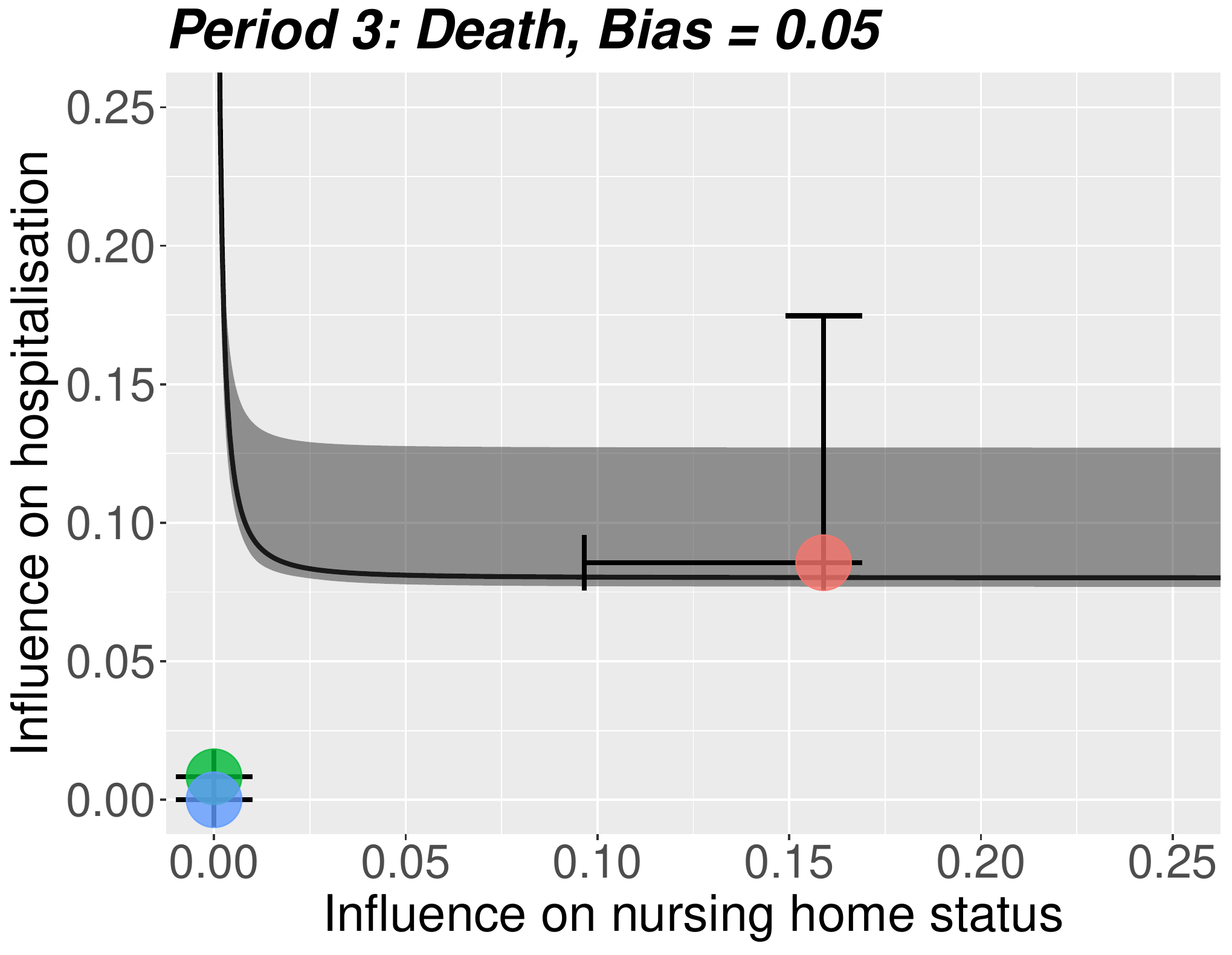}\includegraphics[scale=0.4]{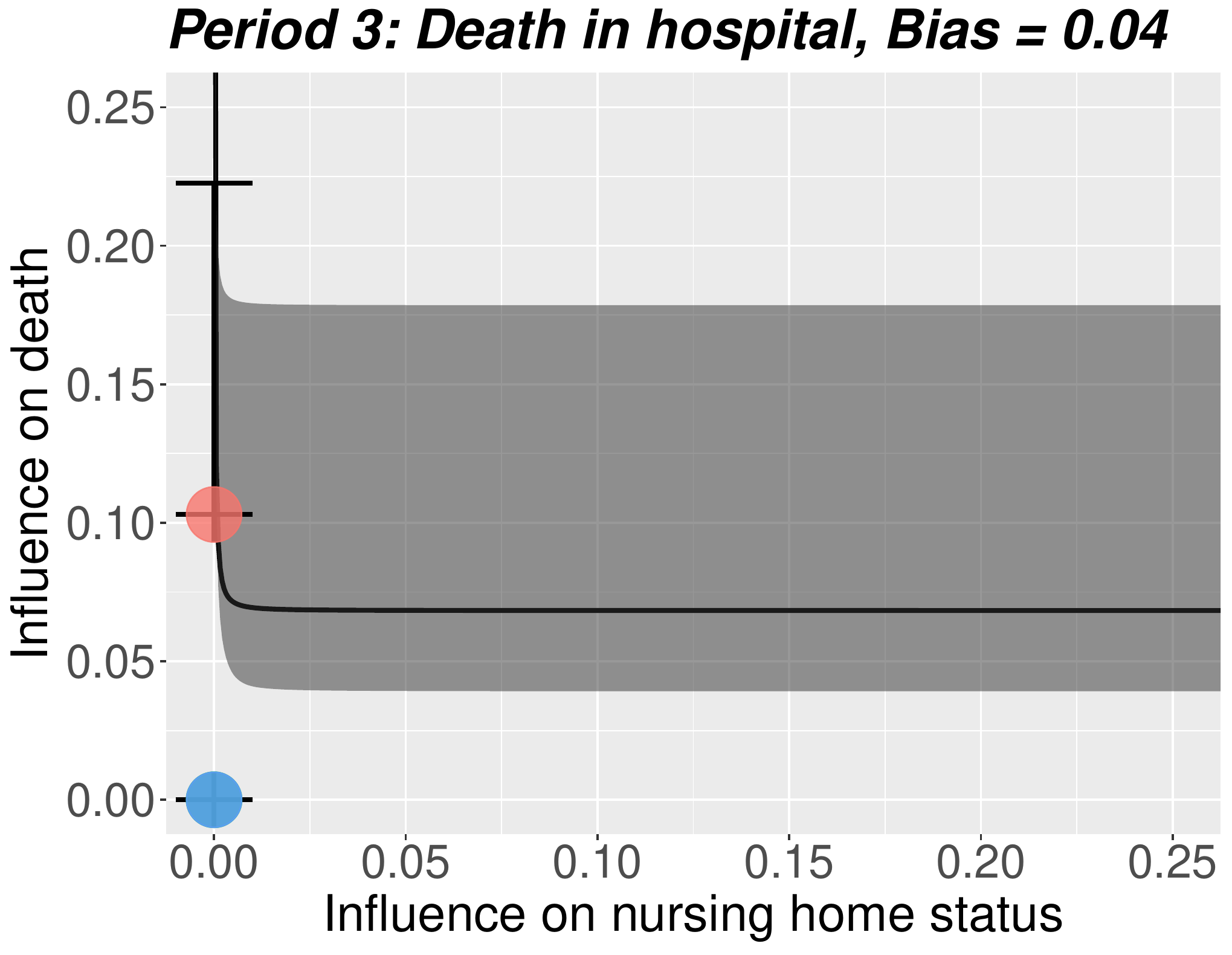}
    \includegraphics[scale=0.2]{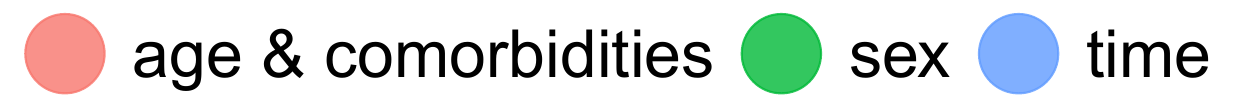}
    \caption{Austen plots for different outcomes, periods, and bias levels. 
    The influence of an unobserved variable that produces the required bias for the original data is shown in solid black and in gray its 95\% bootstrap confidence interval. Points represent the influence of groups of measured covariates for the original data and error bars the corresponding 95\% bootstrap confidence intervals. For the construction of the confidence intervals, 100 bootstrap replicates were taken.}
    \label{fig:austen_plots}
\end{figure}

To study the sensitivity of the results to unobserved variables, we display several Austen plots\cite{veitch2020sense} in Figure \ref{fig:austen_plots}. Details are delayed to the \nameref{sec_Methods} section. The high-level idea is the following. One fixes a level of (statistical) bias produced by the unobserved variable (Bias in the title of the plots) which can make the confidence intervals of half of the estimators contain the 0, hence, blocking a majority decision in favour of unfairness. Based on some model assumptions (details are given in section \nameref{sec_Sensitivity}), one can estimate the influence that this unobserved variable should have on the nursing home status and on the outcome to produce the required amount of bias. This is depicted by the black solid line, for the original data, and the 95\% bootstrap confidence intervals in gray. One can also estimate the influence in  the nursing home status and outcome of groups of measured covariates. This is illustrated as points for the original data, and as error bars for the 95\% bootstrap confidence intervals.

Based on this, top-right of Figure \ref{fig:austen_plots} conveys that to produce enough change on ATE estimates to overturn our conclusion on unfairness in hospitalisation for period 3, an unobserved covariate with the same influence on nursing home status as the combination of age and all used comorbidities should have the same effect on hospitalisation. Likewise, for death in period 3, the required unobserved variable should have a similar influence to that of the combination of age and comorbidities. Notice that the plots suggest that for almost any level of influence of the unobserved variable on the nursing home status, the influence on the outcome should be comparable to that of age and comorbidities combined. 

\subsection*{Quality of ATE estimation}\setcurrentname{Quality of ATE estimation}\label{sec_Quality}
In causal inference, when matching or inverse probability treatment weighting is used, one has to ensure that the resulting (weighted) (sub)samples are behaving appropriately\cite{stuart2010matching, garrido2014methods, nogueira2022methods, zhao2021propensity}. Diagnostics try to confirm that the weighted samples for each class of $Z$ have joint distributions in the covariates $\mathbf{X}$ that are as similar as possible. Notice that in causal inference, the role of covariates $\mathbf{X}$ in the unconfoundedness hypothesis can be played by the propensity score $e(\mathbf{X})=E(Z|\mathbf{X})$\cite{imbens2015causal}. Hence, similarity of the distribution of $e(\mathbf{X})$ for residents and non-residents is also a desirable feature.

\begin{figure}[h!]
    \centering
    \includegraphics[scale=0.29]{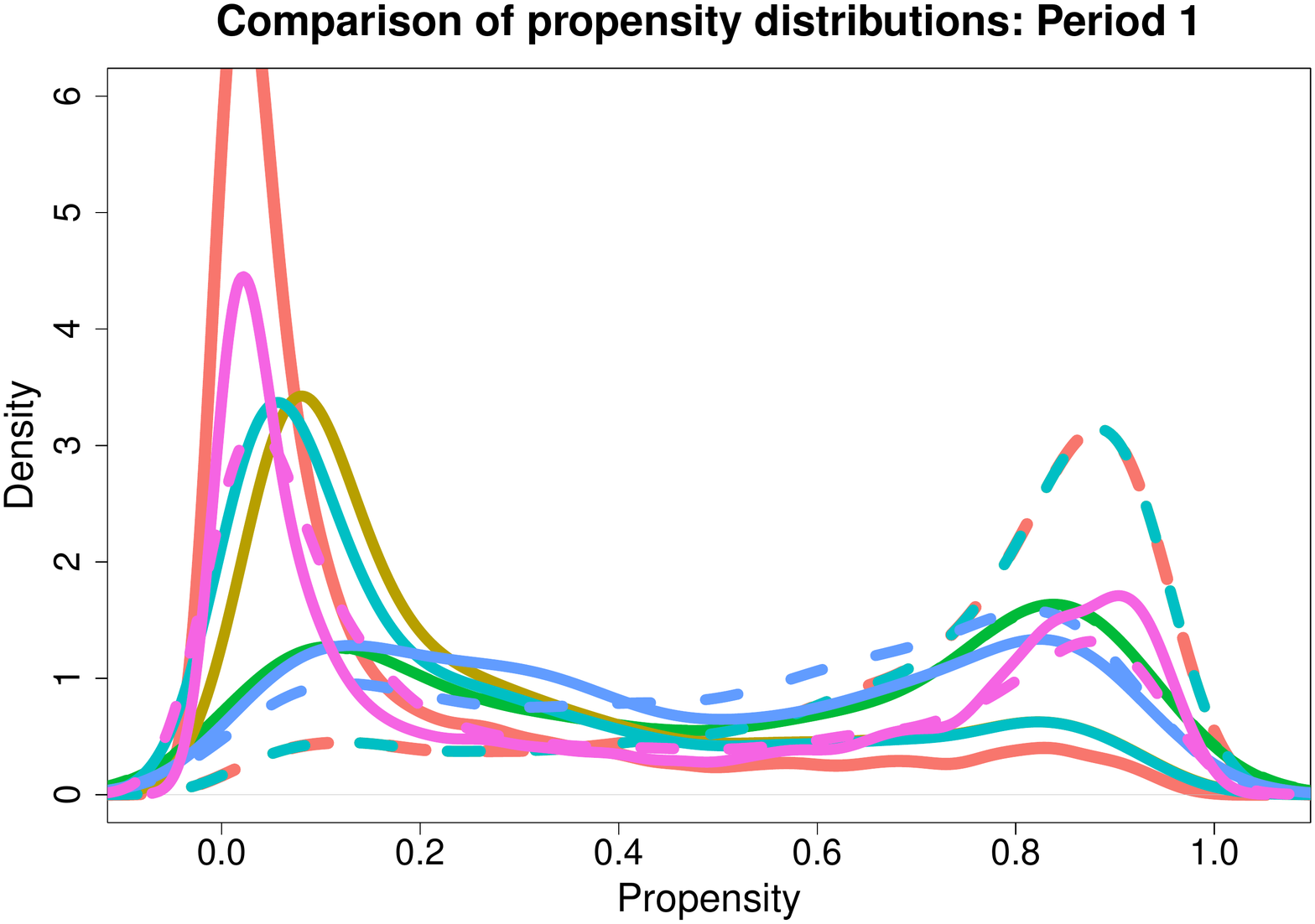}
    \includegraphics[scale=0.29]{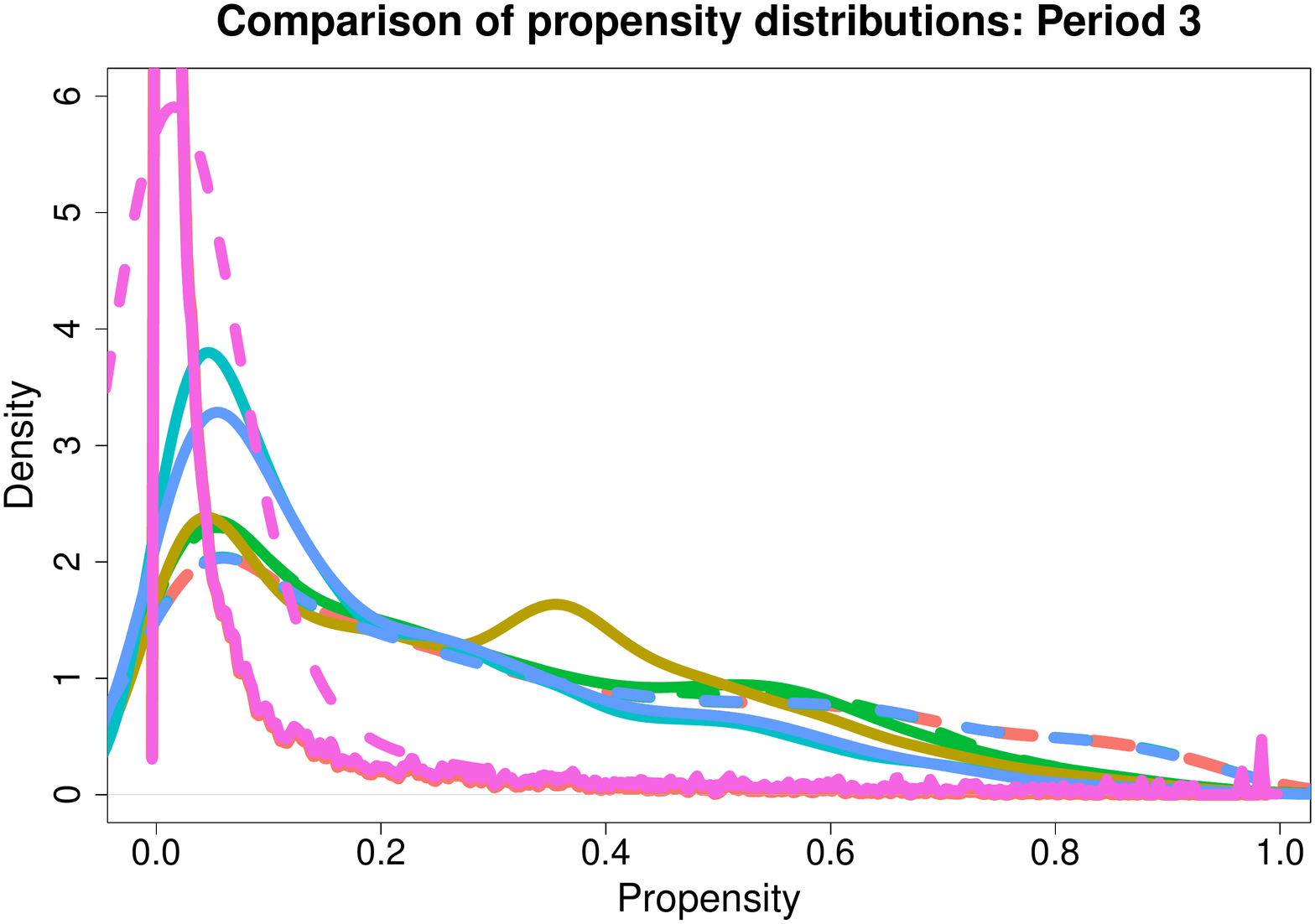}
    \includegraphics[scale = 0.4]{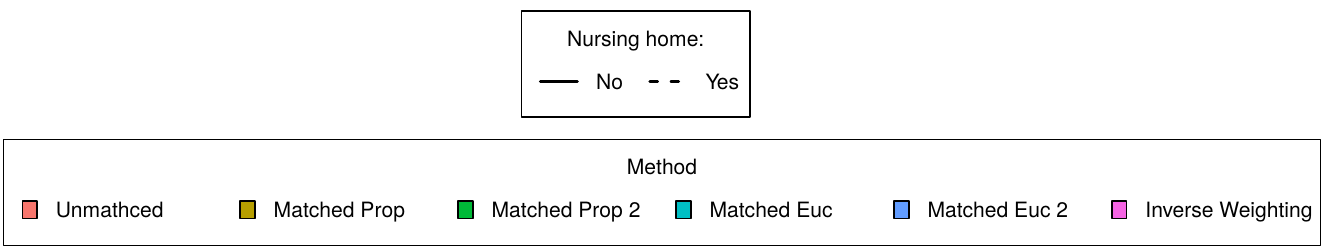}
    \caption{Distribution of propensity score estimates for the data used in \emph{outcomes hospitalisation and death} for nursing home residents, $Z=1$, in dashed, and non-residents, $Z=0$, in solid. Methods correspond with Table \ref{table:characterization_estimations}, where both unmatched methods produce the same propensity scores, which is also the case for the inverse weightings.}
    \label{fig:prop_comparison}
\end{figure}

In Figure \ref{fig:prop_comparison} one can see that all methods reduce the difference between propensity score distributions of nursing home residents and non-residents compared to the original ones in red. Notoriously, Matched Prop 2 in green, which represents a matching method based on distance in propensity score achieves almost perfect equality between the two distributions in both periods. It is worth noticing that in most periods (see Appendix \ref{Quality_ATE_estiamtion}) inverse weighting is trying to make the propensity score distribution of nursing home residents more similar to that of non-residents. By design (see \nameref{sec_Methods}), matching methods try the opposite, to make the propensity score distribution of non-residents more similar to that of nursing home residents. This is easily seen for Period 3 in Figure \ref{fig:prop_comparison}.

Let us stress that since propensity score is an estimated quantity, the similarity between propensity distributions does not guarantee similarity in the original covariates that were used in the methods, $\mathbf{X}$, nor in the whole set of available covariates. However, below we provide some evidence that this is indeed the case.

\begin{figure}[h!]
    \centering
    \includegraphics[scale=0.3]{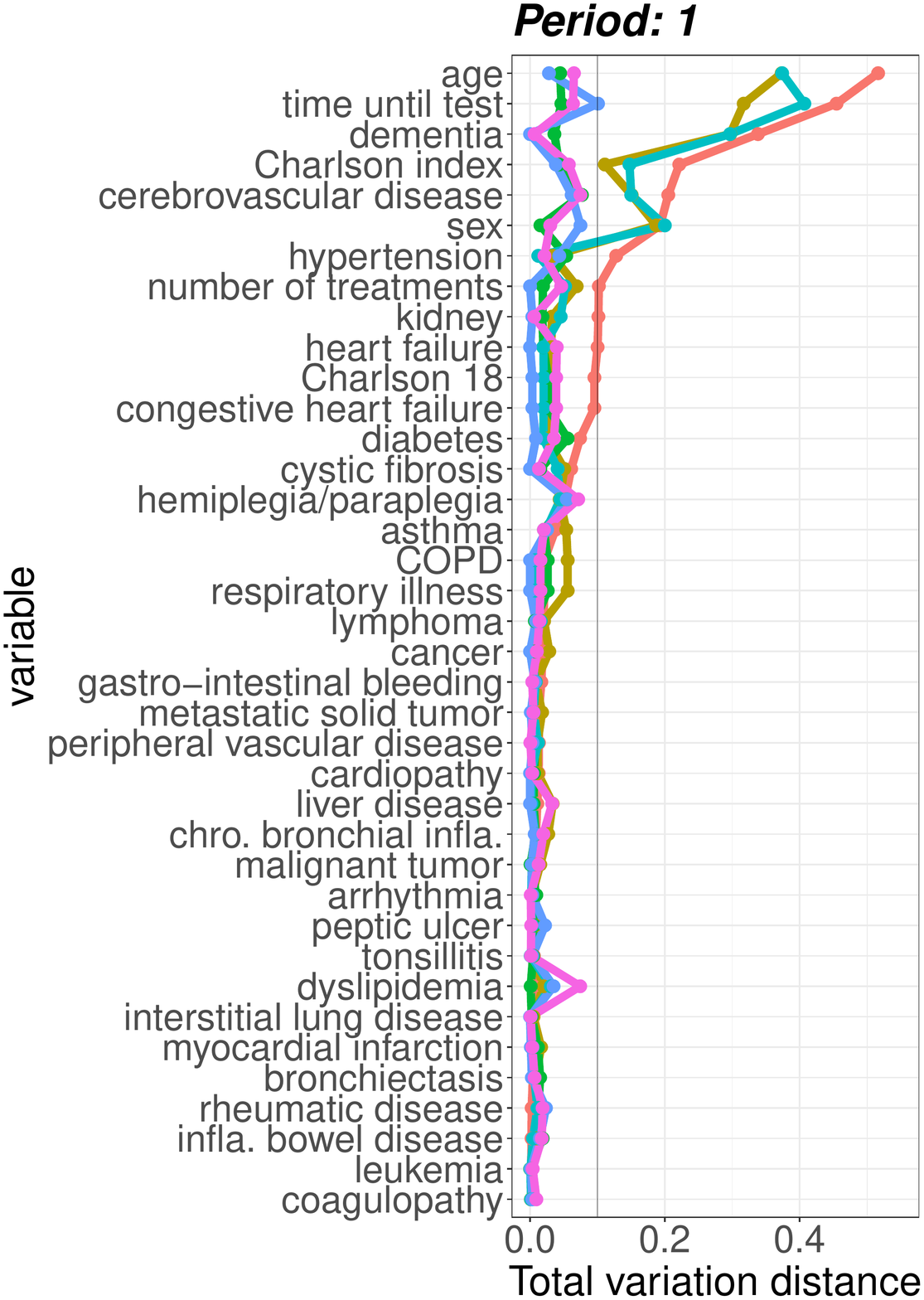}\includegraphics[scale=0.3]{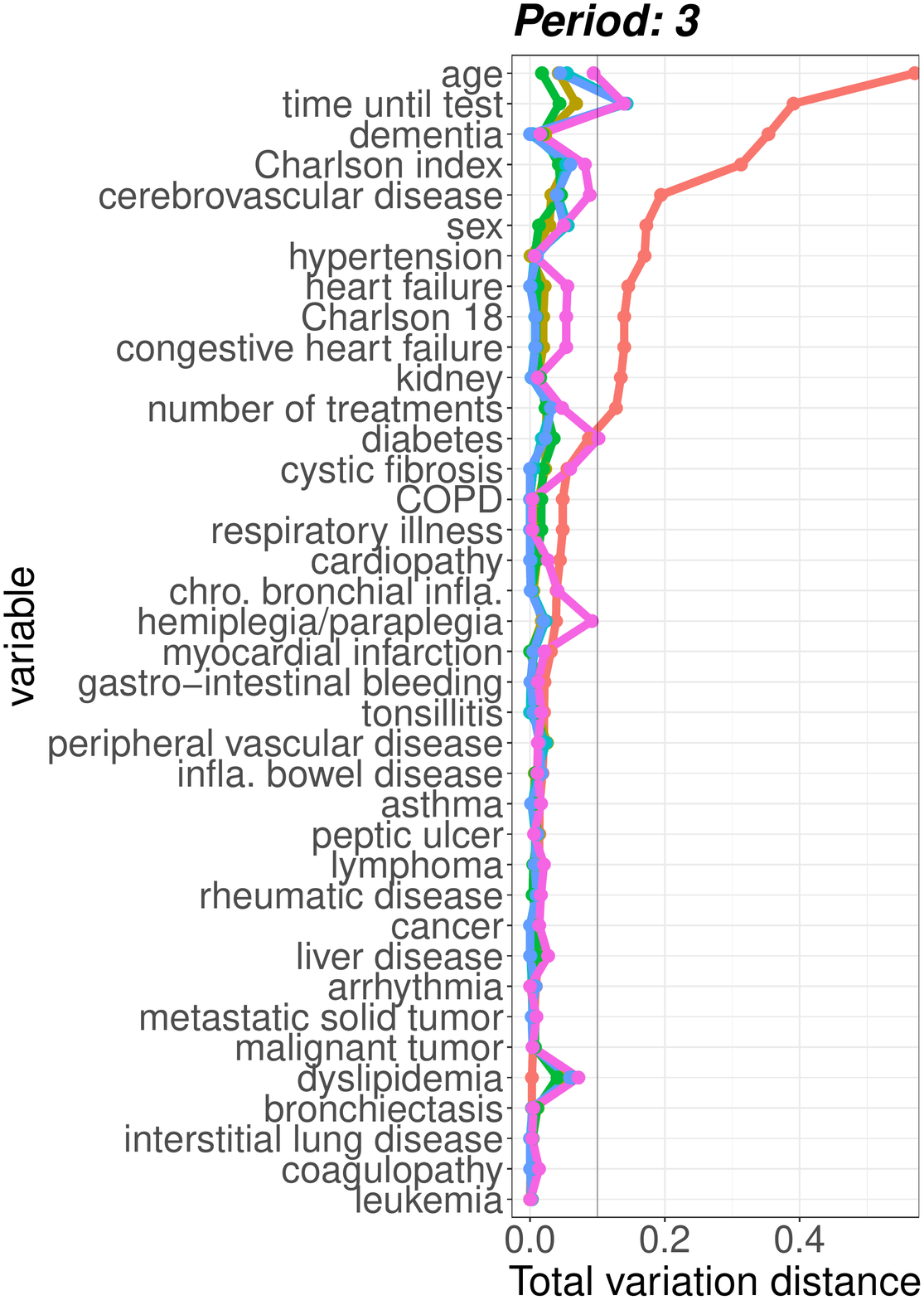}
    \includegraphics[scale = 0.35]{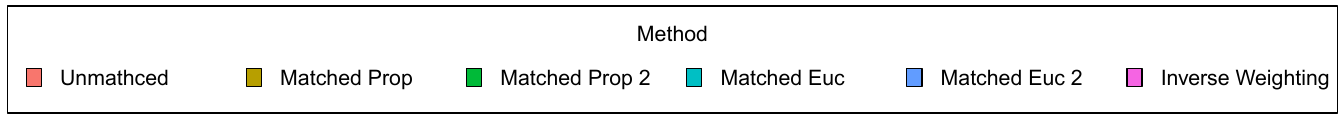}
    \caption{Estimates of the Total Variation distance between the marginals corresponding to nursing home residents and non-residents for the data used in \emph{outcomes hospitalisation and death}.}
    \label{fig:dtv_comparison}
\end{figure}

Figure \ref{fig:dtv_comparison} shows estimates of the distance in total variation for each marginal (for the whole set of available covariates) between nursing home residents and non-residents. Total variation distance between two probability measures $P,Q$ in the same $\sigma$-algebra $\mathcal{A}$ is defined as $d_{TV}(P,Q)=\sup_{A\in\mathcal{A}}|P(A)-Q(A)|$. Informally, it is the maximum difference in probability assigned to the same event. Hence, values close to 0 mean that the probabilities are very similar.
\begin{table}[h!]
\centering
\begin{small}
\begin{tabular}{ccccccccc}
                  & \multicolumn{8}{c}{Multivariate similarity: Hospitalisation, Death}                                                                                                                                                                  \\ \cline{2-9} 
                  & \multicolumn{2}{c}{Period 1}                        & \multicolumn{2}{c}{Period 2}                        & \multicolumn{2}{c}{Period 3}                        & \multicolumn{2}{c}{Period 4}                        \\ \cline{2-9} 
                  & MMD$^2$                  & $\mathcal{W}_2$          & MMD$^2$                  & $\mathcal{W}_2$          & MMD$^2$                  & $\mathcal{W}_2$          & MMD$^2$                  & $\mathcal{W}_2$\\ \cline{2-9} 
Unmatched         & 0.1922                   & 1.4337                   & 0.1895                   & 1.4661                   & 0.2086                   & 1.5383                   & 0.2163                   & 1.6044                   \\ \hline
Matched Euc       & 0.1383                   & 1.2547                   & 0.0539                   & 1.0962                   & 0.0479                   & 1.1507                   & \textit{\textbf{0.0391}}                   & 1.2291 \\ \hline
Matched Euc 2     & 0.0618                   & \textit{\textbf{0.9892}} & 0.0433                   & \textit{\textbf{1.0523}} & 0.0437                & \textit{\textbf{1.0896}}                   & 0.0399                   & 1.1661                   \\ \hline
Matched Prop      & 0.1384                   & 1.3121                   & 0.0410                   & 1.1292                   & 0.0452                   & 1.2038                   & 0.0411                   & 1.2507                   \\ \hline
Matched Prop 2    & \textit{\textbf{0.0521}} & 1.2932                   & \textit{\textbf{0.0309}} &1.1471                    & \textit{\textbf{0.0324}} & 1.2188                   & 0.0472                   & 1.2727                   \\ \hline
Inverse Weighting &                          &  1.2125                  &                          & 1.1171                   &                          & 1.2400                   &     & \textit{\textbf{1.1362}}                         \\ \hline
\end{tabular}
\vspace*{10pt}

\begin{tabular}{ccccccccc}
                  & \multicolumn{8}{c}{Multivariate similarity: Death in hospital}                                                                                                                                                        \\ \cline{2-9} 
                  & \multicolumn{2}{c}{Period 1}                        & \multicolumn{2}{c}{Period 2}                        & \multicolumn{2}{c}{Period 3}                        & \multicolumn{2}{c}{Period 4}                        \\ \cline{2-9} 
                  & MMD$^2$                  & $\mathcal{W}_2$          & MMD$^2$                  & $\mathcal{W}_2$          & MMD$^2$                  & $\mathcal{W}_2$          & MMD$^2$                  & $\mathcal{W}_2$          \\ \cline{2-9} 
Unmatched         & 0.1606                   & 1.5880                   & 0.1323                   & 1.5782                   & 0.1591                   & 1.6803                   & 0.1507                   & 1.9034                   \\ \hline
Matched Euc       & 0.0593                   & 1.3852                   & 0.0494                   & 1.4437                   & 0.0734                   & 1.5432                   & 0.0816                   & 1.7431                   \\ \hline
Matched Euc 2     & \textit{\textbf{0.0542}} & \textit{\textbf{1.3483}} & 0.0590                   & \textit{\textbf{1.4264}} & 0.0796                   & \textit{\textbf{1.5270}} & \textit{\textbf{0.0685}} & \textit{\textbf{1.6637}} \\ \hline
Matched Prop      & 0.0629                   & 1.4596                   & \textit{\textbf{0.0425}} & 1.4841                   & \textit{\textbf{0.0689}} & 1.6006                   & 0.0985                   & 1.8629                   \\ \hline
Matched Prop 2    & 0.0681                   & 1.5283                   & 0.0446                   & 1.4895                   & 0.0772                   & 1.6359                   & 0.1203                   & 1.8543                   \\ \hline
Inverse Weighting &                          & 1.4919                   &                          & 1.4665                   &                          & 1.6563                   &                          & 1.9129                    \\ \hline
\end{tabular}
\end{small}
\caption{Multivariate similarity measured by distances between multivariate probabilities. MMD$^2$ is the square of the Maximum Mean Discrepancy as defined in (\ref{eq:mmd}) computed for the radial kernel. $\mathcal{W}_2$ is an estimate of the 2-Wasserstein distance (\ref{eq:wasser}). More details in the \nameref{sec_Methods}' section \nameref{sec_diagnostics}.}
\label{table:MMD_h0_rejection}
\end{table}
In Figure \ref{fig:dtv_comparison} one can see that for the covariates where the discrepancy was the biggest (highest values of the red curve) a significant reduction is achieved by Inverse Weighting, and by both Matched Euc 2 and Matched Prop 2. Remarkably, substantial reduction is achieved in many of the covariates that were not directly used (the ones that are in the $y$-label of Figure \ref{fig:dtv_comparison} but not in (\ref{eq:covariates})), for example, cerebrovascular disease, hypertension, diabetes, and cystic fibrosis. However, for the covariates where there were almost no discrepancies between the original marginal distributions, some have appeared, e.g., for dyslipidemia. Crucially, these remain low and therefore can be considered tolerable for the ATE estimation.

Nonetheless, similarity in marginals is not necessarily a sign of similarity in the corresponding joint distributions. To tackle this we make use of two different distance measures between multivariate probability distributions. On the one hand, we use Maximum Mean Discrepancy (MMD) which is the base for a multivariate goodness of fit test\cite{gretton2006kernel}. The null hypothesis ($H_0$) is that nursing home residents and non-residents, viewed as two populations, come from the same distribution over all the available covariates. The alternative is that they come from different distributions. Since, in its current form, the estimation procedure does not allow for weights, we can only apply it to the matched samples which do produce different data to the original ones, but without taking weights into account. It has been suggested that it is more meaningful to present the statistics, i.e., the distance (discrepancy) between the distributions, instead of presenting the results of the tests (which can be seen in Appendix \ref{Quality_ATE_estiamtion})\cite{austin2015moving}. The former are shown in Table \ref{table:MMD_h0_rejection}. On the other hand, we use the 2-Wasserstein distance, which comes from the field of optimal transport, and can be applied to weighted samples\cite{peyre2019computational}. Details can be found in \nameref{sec_Methods}. From Table \ref{table:MMD_h0_rejection}, it is clear that, in general, data are more similar in multivariate space after matching or inverse weighting (lower values of the distances) in all periods. Particularly,  Matched Euc 2 and Matched Prop 2 produce the best performance (smallest distance estimates) in many of the cases. 

\section*{Discussion}\setcurrentname{Discussion}\label{sec_Discussion}

Our results show that there is evidence that from the beginning of the COVID-19 pandemic in the Basque Country (1/3/2020) until more than a year and a half later (13/12/2021) nursing home status directly affected outcomes as hospitalisation, mortality, and in-hospital mortality in different degrees. In other words, there is evidence that nursing home residents had worse outcomes due to, at least in part, being in a nursing home. This begs a difficult question: Is this something to be expected, or is it a sign of systematic discrimination? If one takes a nursing home to be simply a type of place where people live, a direct effect of that magnitude on the studied outcomes seems unwarranted. Nevertheless, nursing home status is more frequently a proxy for some individuals characteristics (related to high level of frailty, lack of independence, or consciousness, or a mental status affection) not available for this study and in that case, its influence on the outcomes may be justified. Yet, as shown in Figure \ref{fig:austen_plots}, such unmeasured frailty variables should have quite strong effects on the outcomes themselves. Further deliberation on the possible effects of relevant unmeasured variables from a medical perspective is provided at the end of this section. The next few paragraphs are concerned with some key statistical issues. 

An important point is that the nursing home status effect is noticeable using two different approaches to obtain comparable sub-populations. One is based on inverse propensity score weighting, where nursing home residents that are very similar to prototypical non-residents are mainly considered. Another one uses trimmed optimal matching and its focus is the other way around, on non-residents whose distribution matches closely the one of nursing home residents. The persistence of the effect suggests that our results are not mere data selection artefacts. An interesting case is hospitalisation during periods 2 and 3. For these, inverse weighting methods do not provide evidence of unfairness, while trimmed matching methods do. Additionally, the adjusted formula method (Unmatched 2) provides evidence of unfairness but of lower magnitude. A possible interpretation is that if one focuses mainly on nursing home residents that correspond better with non-residents (younger and healthier), there is small or no unfairness regarding hospitalisation. However, if one selects a subgroup of nursing home non-residents that matches better the whole variety of nursing home residents' profiles, a strong unfairness effect is observed.

The quality diagnostics, especially the one\textcolor{olive}{s} shown in Table \ref{table:MMD_h0_rejection}, confirm that the obtained sub-populations are in fact closer, essentially in the Euclidean space of all the measured covariates, than the original ones. This is an indication of methods working properly, and therefore fulfilling their purpose, which can give us more confidence in the conclusions we have drawn.

As we are dealing with sensitive issues, we want to highlight some limitations of our approach. The first one is well known, causal inference relies on strong assumptions. We have tried to partially address the unconfoundedness assumption, but we did not challenge the other ones (positivity, consistency, and no interference); as far as we know it is not immediate how to do that. A related issue is that of residual confounding due to the fact that the joint distributions of nursing home residents and non-residents are not the same, even after applying corrections. This means that some or all of the observed effects can be caused by this difference. Although the latter seems unlikely in view of the diagnostics of section \nameref{sec_Quality}, there are no statistical guarantees. Producing such guarantees or bounds on the ATE estimates purely due to differences in the joint probabilities is an exciting challenge that can have huge applications in the field of causal inference. Another point of possible improvement to be explored concerns the chosen similarity criterion. Although the Wasserstein distance has proven to be a very good solution for measuring closeness in Euclidean space, perhaps such a goal is not the most appropriate in this context. This concern can be addressed since the $d$-Wasserstein distance is just an optimal transport cost, where the cost function is a $d$-power of the Euclidean distance. However, a more suitable cost function, if available, could be used.

Once limitations are accounted for, we believe that the main takeaway point of our work is the rigorous statistical analysis of a controversial issue that is in the public spotlight. It seems clear that further study and data collection is required to reach more definitive conclusions. However, our work does show evidence of the concerning possibility of nursing home residents being systematically on the bad end for the studied outcomes. Especially concerning are our results for the beginning of the COVID-19 pandemic (period 1). Here, our analysis provides compelling evidence of unfairness to nursing home residents in terms of hospitalisation and mortality, and points to the vulnerable situation of these individuals in extraordinarily grim circumstances. The first stages of the COVID-19 pandemic, not only in Spain but also in many other developed countries, greatly affected individuals who were older, with severe comorbidities, and who were residing in nursing homes. This could be related to the explosive start of the pandemic, the total saturation of the primary care services, emergency services, and hospital beds, and the lack of proper material to establish diagnosis, protection, and isolation measures or effective treatments \cite{Grimm2021hospital, costa2021fatal, miralles2021unmet, morciano2021excess}. This had heavy consequences. Particularly in what we have considered period 1, high rates of mortality and non corresponding low rates of hospital admission among nursing home residents were observed. 

Next, we provide some plausible causes for the observed statistical unfairness in the studied outcomes. In the Basque Country it has been reported that the risk of dying from COVID-19 in nursing homes depended fundamentally on the individual characteristics of the patient (age, sex, cognitive status, and functional status) \cite{Ararteko:2021}. This is in concordance with the results of Mehta \emph{et al.} (2021) \cite{mehta2021risk} in what is probably the most comprehensive study (with more than 480,000 people) published to date on the factors associated with COVID-19 infection and mortality in nursing homes. In the same line, the study by Suñer \textit{et al.} (2021)\cite{suner2021retrospective} in 167 nursing homes in Catalonia (autonomous community in Spain) reported that a 10\% increase in the proportion of residents with complex chronic disease increased the risk of mortality by 7\%. Also, it is important to stress that nursing home residents live in an environment where there is a higher risk of infection at any time due to contact with far more people than similar individuals living on their own. Such higher risk of infection among nursing home residents could have affected the ATE estimates provided in this work.

Some other particularities in relation to nursing home residents must be considered. In the Basque Country, the explosive situation at the beginning of the pandemic led to the opening of other forms of health care for these patients who were first attended through home hospitalisation services and later in nursing homes equipped for health care. This protocol was in place until the end of 2021 (comprising periods 1, 2 and 3 in this study), close to the beginning of the period dominated by the Omicron variant (period 4). Hence, this could be (partially) responsible for the negative ATE estimates we have observed for hospitalisation, as the number of admissions to regular hospital centers for periods 1, 2 and 3 was reduced in nursing home residents. Additionally, restriction measures to contain the infection spread were applied more strictly in nursing homes, which had important side effects \cite{hugelius2021consequences}. Isolation in this situation is especially harmful due to physical and spatial limitations (confinement to the bedroom), and these limitations prevent individuals from adapting to the new situation, resulting in a worsening of dementia and other mental health conditions.

To conclude, algorithmic fairness is a mathematical definition of what is considered (statistically) fair (just, equitable), and one should bear in mind that it is loosely based on ethical but definitely not on any clinical considerations. The unprecedented COVID-19 crisis is a powerful reason to reconsider the best way of managing care for the sick elderly: we should recognize the particular characteristics, needs, and risks of the elderly in our protocols and health planning \cite{aronson2020age}.

\section*{Methods}\setcurrentname{Methods}\label{sec_Methods}
In this section, we supply details on the data and methods used in this work. In \nameref{sec_Data} the particularities of data extraction, ethic statements, and variable's definition are delineated. In \nameref{sec_Procedures}, we expand on the ATE estimators summarised in Table \ref{table:characterization_estimations}, on the model to produce the Austen plots of Figure \ref{fig:austen_plots} and on some of the methods used in the diagnostics of the ATE estimators.  
\subsection*{Data}\setcurrentname{Data}\label{sec_Data}
All patients included in this retrospective cohort study were residents of the Basque Country who had a SARS-CoV-2 infection, laboratory-confirmed by a positive result on the reverse transcriptase-polymerase chain reaction assay for severe acute respiratory syndrome coronavirus 2 (SARS-CoV-2) or a positive antigen test between March 1, 2020, and January 9, 2022. From March 1, 2020, to July 31, 2020, positive IgM or IgG antibody tests performed due to patients having symptoms suggestive of the disease or having had contact with a positive case were also included in the general population sample. The first positive from each patient was collected. Only patients aged over 18 years were included. Data identifying people nursing home status were obtained from the Basque Health Department. However, in this work, only individuals strictly older than 59 were considered since this age group represented more than 99.9\% of the population in nursing homes. The study protocol was approved by the Ethics Committee of our area (reference PI2020123). All patient data were kept confidential.
Data collected include sociodemographic, baseline comorbidities, 
baseline treatments, 
dates of hospital admission and discharge and whether patients were admitted to an intensive care unit (ICU), and vital status. A detailed explanation of the data collected is given elsewhere \cite{portuondo2023clinical, espana2023impact}.

The outcomes used in the study were as follows: 1) Hospital admission due to COVID-19, defined if admission occurred within 15 days of the patient’s testing positive, when the positive test preceded hospitalisation, and up to 21 days after admission when the patient tested positive during hospitalisation; 2) Death during the three months following diagnosis or three months from discharge; and, 3) Death during hospital admission. A complete descriptive table of the data is available in Appendix \ref{descriptive_analysis}.

For each period the following variables were defined:
\begin{small}
    \begin{verbatim}
    time until test = Date positive - Date first positive in the period
    time until hosp = Date hospitalisation - Date first positive in the period
    time positive to hosp = time to hosp - time to test.
\end{verbatim}
\end{small}
The rationale behind these variables is the following. `time until (SARS-COV-2 positive) test' and `time until hosp' are considered proxies for the state of the public health care system at the time, i.e., if two individuals with a different nursing home status test positive or are hospitalised close in time then the public health care system situation is expected to be similar.  `time positive to hosp' is a proxy of the possible severity of the disease at the moment of hospitalisation, presumably, similar values will correspond to individuals with similar severity.  

\subsection*{Procedures}\setcurrentname{Procedures}\label{sec_Procedures}
Causal inference is concerned with the behaviour of the potential outcomes $Y(1), Y(0)$, the outcome when treated (in our context, when resident in a nursing home) and the outcome when non-treated (non-resident in a nursing home), respectively. In this setting, $\text{ATE}=E(Y(1)-Y(0))$. However, in reality, one only observes the outcome for one of the instances, either treated or non, and does not have access to the counterfactual. Estimation of $ATE$ becomes possible under some strong assumptions. These allow to write the observed outcome as $Y=ZY(1)+(1-Z)Y(0)$. From the unconfoundedness assumption, $(Y(1),Y(0))\perp Z|\mathbf{X}$, it follows that $(Y(1),Y(0))\perp Z|e(\mathbf{X})=E(Z|\mathbf{X})$, which means that potential outcomes are unaffected by treatment assignment if they correspond to the same propensity score\cite{lunceford2004stratification}. This is the main reason for propensity score's popularity, it provides a strong dimension reduction, from the dimension of $\mathbf{X}$ to one.

Returning to the estimation of ATE, it can be done in many ways, and there is extensive literature on the topic\cite{rubin_2006, imbens2015causal, lunceford2004stratification, neal2020introduction, stuart2010matching}. In this regard, we propose a novel way of estimating ATE based on matching samples by trimmed optimal transport. Additionally, we propose some modifications to the standard procedures of diagnostics\cite{austin2015moving}, that may help robustify the endeavour and produce firmer evidence. Details are provided below.  
\subsubsection*{ATE estimators}
It is well known that in a randomised study association of treatment and outcome equates causation. This means that
\begin{equation}
\label{ATE_est}
\text{ATE} = P(Y=1|Z=1) - P(Y=1|Z=0), 
\end{equation}
and estimating ATE is just a matter of estimating the outcome probability in each treatment class, in our case nursing home residents and non-residents. Let our sample be $\{(Y_i,Z_i,\mathbf{X}_i)\}_{i=1}^n$. Then the estimator we have called Unmatched in Table \ref{table:characterization_estimations} is defined as
$$\mathrm{ATE}_{\mathrm{Unmatched}}= \frac{\sum_{i=1}^nZ_iY_i}{\sum_{i=1}^nZ_i}-\frac{\sum_{i=1}^n(1-Z_i)Y_i}{n -\sum_{i=1}^nZ_i}.$$

However, the data at hand correspond to an observational study, hence ATE estimation should be adapted to this case. The estimator Unmatched 2 uses equation (\ref{eq:fairness_ate}), also known as adjustment formula,  and produces an estimate of the conditional expectation $E(Y|Z,\mathbf{X})$, which we denote $\hat{E}(Y|Z,\mathbf{X})$ and evaluations of the estimator on a (possibly counterfactual) data point $(Z_0,\mathbf{X_i})$ are denoted as $\hat{E}(Y|Z=Z_0,\mathbf{X}_i)$. Then the estimate we used is
$$\mathrm{ATE}_{\mathrm{Unmatched\,2}}=\frac{1}{n}\sum_{i=1}^n\left(\hat{E}(Y|Z=1,\mathbf{X}_i) - \hat{E}(Y|Z=0,\mathbf{X}_i)\right).$$
$\hat{E}(Y|Z,\mathbf{X})$ can be obtained by any quadratic loss optimisation procedure. When the outcome was hospitalisation and death we used gradient boosting methods. When it was death in hospital we used random forests. We selected the method that produced the best loss function from three candidates: logistic regression, gradient boosting, and random forest.

Inverse propensity score (aka, probability treatment) weighting (IPSW) relies on the estimation of the propensity score $E(Z|\mathbf{X})$ through an estimator $\hat{E}(Z|\mathbf{X})$. As it is a conditional expectation, we can proceed as in the previous paragraph, selecting the best performing method. Again, we used gradient boosting for the general data, and random forest when we looked only at hospitalised individuals. One can produce different estimates of ATE with the help of propensity scores\cite{lunceford2004stratification, austin2011introduction, neal2020introduction}. The ones we chose were the following;
$$\mathrm{ATE}_{\mathrm{Inverse\,Weighting}}=\frac{1}{n}\sum_{i=1}^n\left(\hat{E}(Y|Z=1,\mathbf{X}_i) - \hat{E}(Y|Z=0,\mathbf{X}_i) + \frac{Z_i(Y_i-\hat{E}(Y|Z=1,\mathbf{X}_i))}{\hat{E}(Z=1|\mathbf{X}_i)}- \frac{(1 -Z_i)(Y_i-\hat{E}(Y|Z=0,\mathbf{X}_i))}{1-\hat{E}(Z=1|\mathbf{X}_i)}\right),$$
$$\mathrm{ATE}_{\mathrm{Inverse\,Weighting\,2}}=\frac{1}{n}\sum_{i=1}^n\left(\frac{Z_iY_i}{\hat{E}(Z=1|\mathbf{X}_i)}- \frac{(1 -Z_i)Y_i}{1-\hat{E}(Z=1|\mathbf{X}_i)}\right).$$
Notice that the estimator Inverse Weighting is what is known as doubly robust.

We now turn our attention to a second set of ATE estimators which are based on first producing samples from the two classes, nursing home residents and non-residents (treated and non-treated), that are as similar as possible in $\mathbf{X}$, and therefore in their joint distributions; and then using (\ref{ATE_est}) to estimate the ATE based on the new samples. In this setting, a popular strategy to achieve the first goal is to try to match both samples. Matching refers to an assignment of members from one group to members from the other in a way that they are similar in some sense. For example, in a one-to-one matching, one extracts a subgroup of the majority class which is the most similar in some sense to the minority class. In some sense, usually means with respect to a distance measure. Selecting an adequate distance can be a major issue. In our study, we used one dimensional distance in propensity score and a (weighted) multivariate Euclidean distance between the normalised data (Propensity Score and Euclidean in Table \ref{table:characterization_estimations}, respectively). Normalisation consisted in bringing all individual variables within a maximum distance of 1, which was done by dividing by the maximum observed distance in the original variable.

Once a distance is chosen, the matching method has to be selected. Once more, the choice of method is important\cite{stuart2010matching}. Here we propose to use a very flexible matching mechanism. The idea is to allow for parts of one or both samples to be down weighted and then matched. The objective is to select the core of both samples that are the most similar and then estimate ATE on the new data using  (\ref{ATE_est}). The methods followed here are based on trimmed comparison of samples and trimmed optimal transport\cite{alvarez2008trimmed, agullo2018trimming, gordaliza2022making}. Let $\{\mathbf{X}^0_i\}_{i=1}^{n_0}$, $\{\mathbf{X}^1_j\}_{j=1}^{n_1}$ be the samples of the covariates corresponding to the non-residents and residents, respectively. We want to solve the following trimmed transportation problem
\begin{equation}
\label{eq:transport_problem}
	\begin{array}{ll@{}ll}
&\min_{\Pi} \sum_{i=1}^{n_0+1}\sum_{j=1}^{n_1+1}\Pi_{ij}\tilde{C}_{ij}&\\
		\text{subject to} 	& \sum_{j=1}^{n_1+1}\Pi_{ij} =\frac{1/n_0}{1-\alpha_0}, & 1\leq i\leq n_0&\\
		& \sum_{j=1}^{n_1+1}\Pi_{(n_0+1)j} = \frac{\alpha_1}{1-\alpha_1}, & &\\
		& \sum_{i = 1}^{n_0+1}\Pi_{ij} = \frac{1/n_1}{1-\alpha_1} & 1\leq j\leq n_1&\\
		& \sum_{i=1}^{n_0+1}\Pi_{i(n_1+1)} = \frac{\alpha_0}{1-\alpha_0}, & &\\
		& \sum_{j=1}^{n_1}\sum_{i=1}^{n_0+1}\Pi_{ij}=1\\
		 &\Pi_{ij} \geq 0,  &1\leq i\leq n_0+1,1\leq j\leq n_1+1&
	\end{array}
\end{equation}
over the discrete joint probability distributions (transport plans) $\Pi$, $(n_0+1)\times (n_1+1)$ matrices, where $\Pi_{ij}$ is the amount of probability transported between $X^0_i$ to $X^1_j$ and $\tilde{C}$ is a $(n_0+1)\times (n_1+1)$ extended cost matrix as in page 50 in Agull{\'o} (2018)\cite{agullo2018trimming}. The transport problem (\ref{eq:transport_problem}) can be approximately solved efficiently\cite{peyre2019computational, transport}. When the cost function is based on propensity 
\begin{table}[h!]
\centering
\begin{tabular}{lcccccccc}
                                   & \multicolumn{8}{c}{Trimming level: Hospitalisation, Death}                                                                \\ \cline{2-9} 
                                   & \multicolumn{2}{c}{Period 1} & \multicolumn{2}{c}{Period 2} & \multicolumn{2}{c}{Period 3} & \multicolumn{2}{c}{Period 4} \\ \cline{2-9} 
                                   & \multicolumn{8}{c}{Nursing home}                                                                                          \\ \cline{2-9} 
                                   & No            & Yes          & No            & Yes          & No            & Yes          & No            & Yes          \\ \cline{2-9} 
\multicolumn{1}{c}{Matched Euc}    & 42.34\%       & 0.00\%       & 87.91\%       & 0.00\%       & 95.30\%       & 0.00\%       & 95.90\%       & 0.00\%       \\ \hline
\multicolumn{1}{c}{Matched Euc 2}  & 85.00\%       & 75.00\%      & 90.99\%       & 7.50\%       & 96.25\%       & 7.50\%       & 96.86\%       & 7.50\%       \\ \hline
\multicolumn{1}{c}{Matched Prop}   & 42.34\%       & 0.00\%       & 87.91\%       & 0.00\%       & 95.30\%       & 0.00\%       & 95.90\%       & 0.00\%       \\ \hline
\multicolumn{1}{c}{Matched Prop 2} & 70.00\%       & 60.00\%      & 90.99\%       & 7.50\%       & 96.25\%       & 7.50\%       & 96.86\%       & 7.50\%       \\ \hline
\end{tabular}
\vspace*{10pt}

\begin{tabular}{lcccccccc}
                                   & \multicolumn{8}{c}{Trimming level: Death in hospital}                                                                     \\ \cline{2-9} 
                                   & \multicolumn{2}{c}{Period 1} & \multicolumn{2}{c}{Period 2} & \multicolumn{2}{c}{Period 3} & \multicolumn{2}{c}{Period 4} \\ \cline{2-9} 
                                   & \multicolumn{8}{c}{Nursing home}                                                                                          \\ \cline{2-9} 
                                   & No            & Yes          & No            & Yes          & No            & Yes          & No            & Yes          \\ \cline{2-9} 
\multicolumn{1}{c}{Matched Euc}    & 81.26\%       & 0.00\%       & 88.36\%       & 0.00\%       & 95.35\%       & 0.00\%       & 92.67\%       & 0.00\%       \\ \hline
\multicolumn{1}{c}{Matched Euc 2}  & 87.36\%       & 30.00\%      & 97.20\%       & 7.50\%       & 96.30\%       & 7.50\%       & 93.60\%       & 7.50\%       \\ \hline
\multicolumn{1}{c}{Matched Prop}   & 81.26\%       & 0.00\%       & 88.36\%       & 0.00\%       & 95.35\%       & 0.00\%       & 92.67\%       & 0.00\%       \\ \hline
\multicolumn{1}{c}{Matched Prop 2} & 87.36\%       & 30.00\%      & 89.24\%       & 7.50\%       & 96.30\%       & 7.50\%       & 93.60\%       & 7.50\%       \\ \hline
\end{tabular}
\caption{Amount of trimming (down weighting) allowed for when using matching based on the trimmed transportation problem. Methods labelled with a 2 allow for extra trimming in the majority class and for some trimming in the minority class.}
\label{table:trimming_levels}
\end{table}
score estimates, Matched Prop entries in Table \ref{table:characterization_estimations} and \ref{table:trimming_levels}, one has $\tilde{C}_{ij}=\left|\hat{E}(Z|\mathbf{X}^0_{i}) - \hat{E}(Z|\mathbf{X}^1_{j})\right|$ for $1\leq i\leq n_0$, $1\leq j\leq n_1$. For the corresponding Matched Euc entries, one has $\tilde{C}_{ij}=\|\mathbf{w}\odot(\mathbf{X}^0_{i} - \mathbf{X}^1_{j})\|^2$, for $\mathbf{w}$ a weighting which may favour the contributions of some covariates over others and $\odot$ denoting Hadamard product. The weighting was a vector of ones (no weighting) but for the output death in hospital (details are provided in Appendix \ref{Quality_ATE_estiamtion}). The levels of down weighting that were allowed are shown in Table \ref{table:trimming_levels}. Since for periods 2-4 non-residents' sample sizes were very superior, the amount of partial trimming allowed for them was very high. This allows to extract a small core of non-residents that were very similar to the corresponding sample of nursing home residents as shown in Figure \ref{fig:prop_comparison} and \ref{fig:dtv_comparison} and Table \ref{table:MMD_h0_rejection}.

Lastly, confidence intervals were computed using the standard procedure for differences of proportions for $\mathrm{ATE}_{\mathrm{Unmatched}}$, i.e., using asymptotic normality. An appropriate weighted version was used for the four estimators obtained by matching. For the rest, an asymptotic normality approximation was also used.
\subsubsection*{Sensitivity estimates}\setcurrentname{Sensitivity estimates}\label{sec_Sensitivity}
To produce sensitivity estimates one needs to adopt a model. Notice that one seeks to infer something about (unobserved) missing information, and therefore some extra assumptions should be tolerated. Let $U$ be an unobserved (missing) covariate. Take $\tilde{e}(\mathbf{X},U)=E(Z|\mathbf{X},U)$ to be the true unobserved propensity score. Then with the notation of our work the sensitivity model to produce the Austen plots in Figure \ref{fig:austen_plots} is\cite{veitch2020sense}
\begin{align*}
    \tilde{e}(\mathbf{X},U)|\mathbf{X}&\sim \mathrm{Beta}\left(e(\mathbf{X})\frac{1-\eta}{\eta},(1-e(\mathbf{X}))\frac{1-\eta}{\eta}\right)\\
    Z|\mathbf{X},U&\sim \mathrm{Bern}(\tilde{e}(\mathbf{X},U))\\
    E(Y|Z,\mathbf{X},U)&=E(Y|Z,\mathbf{X})+\delta\left(\mathrm{logit}\,\tilde{e}(\mathbf{X},U)-E(\mathrm{logit}\,\tilde{e}(\mathbf{X},U)|Z,\mathbf{X})\right).
\end{align*}
Where $\eta$ is the free parameter setting the influence on nursing home assignment, and $\delta$ the one determining the influence on the outcome. Under this model, $E(E(Z|\mathbf{X},U)|\mathbf{X})=e(\mathbf{X})$, $E(E(Y|Z,\mathbf{X}, U)|Z,\mathbf{X})=E(Y|Z,\mathbf{X})$, and $$bias=\left|\mathrm{ATE} - E_{\mathbf{X}}\left(E(Y|Z=1,\mathbf{X})-E(Y|Z=0,\mathbf{X})\right)\right|=\left|\delta\left(E(\mathrm{logit}\,\tilde{e}(\mathbf{X},U)|Z=1,\mathbf{X})-E(\mathrm{logit}\,\tilde{e}(\mathbf{X},U)|Z=0,\mathbf{X})\right)\right|.$$ More details and estimation strategies are provided in Veitch \textit{et al.} (2020)\cite{veitch2020sense}.
\subsubsection*{ATE diagnostics}\setcurrentname{ATE diagnostics}\label{sec_diagnostics}
To evaluate the quality of the samples obtained by IPSW or by matching one has to compare the joint probability distributions of the covariates for both classes of $Z$. Typically, a quite intuitive first approach is to look at some statistic of the one dimensional distributions over all covariates. The most common statistic is standardised difference, however, it is not well suited for multimodal distributions as is the case of our study. A sound alternative is to use Kolmogorov distance, the supremum norm between the empirical cumulative distributions. Here, we opted for an estimate of the total variation distance which has a very clear and intuitive interpretation. Notice that in 1d one can estimate the $d_{TV}$ by density estimation and numerical integration, for continuous covariates, and, by direct computation for discrete ones. This kind of procedure results in comparisons as the one depicted in Figure \ref{fig:dtv_comparison}. However, very different joint probabilities may have the same marginal distributions, therefore this is not enough to evaluate the similarity between the resulting samples.

To address the full multivariate problem we use distance measures designed for handling multivariate probability distributions and which can be efficiently estimated. Let $P,Q$ be two probability distributions on the same Euclidean space, then the $d-$Wasserstein distance is defined as
\begin{equation}
\label{eq:wasser}
    \mathcal{W}_d^d(P,Q) = \inf_{(\mathbf{S},\mathbf{T}):\mathbf{S}\sim P,\mathbf{T}\sim Q} E\|\mathbf{S}-\mathbf{T}\|^d.
\end{equation}
Notice that the infimum is over all joint distributions with first marginal $P$ and second marginal $Q$. Let $\mathcal{F}$ be a class of functions, then the Maximum Mean Discrepancy is defined as
\begin{equation*}
    \mathrm{MMD}(\mathcal{F},P,Q)=\sup_{f\in\mathcal{F}}\left(E_{\mathbf{S}\sim P}f(\mathbf{S}) - E_{\mathbf{T}\sim Q}f(\mathbf{T})\right).
\end{equation*}
When $\mathcal{F}$ is a subset of a reproducing kernel Hilbert space (RKHS) with kernel $\kappa(\cdot,\cdot)$ one has
\begin{equation}
\label{eq:mmd}
    \mathrm{MMD}^2(\mathcal{F}, P,Q)=E_{\mathbf{S},\mathbf{S}'\sim P}\kappa(\mathbf{S},\mathbf{S}') -2E_{\mathbf{S}\sim P, \mathbf{T}\sim Q}\kappa(\mathbf{S},\mathbf{T})+E_{\mathbf{T},\mathbf{T}'\sim P}\kappa(\mathbf{T},\mathbf{T}') 
\end{equation}
where $\mathbf{S},\mathbf{S}', \mathbf{T},\mathbf{T}'$ are independent between each other. Sample estimates for both (\ref{eq:wasser}) and (\ref{eq:mmd}) exist and can be computed with current software\cite{gretton2006kernel, peyre2019computational, transport, kernlab}. Table \ref{table:MMD_h0_rejection} shows the change in the distance between the multivariate joint distributions for $Z=0$ and $Z=1$ for the different samples. For the MMD computation we used the radial kernel $\kappa_\sigma(\mathbf{s},\mathbf{t})=\mathrm{exp}\left(-\sigma\|\mathbf{s}-\mathbf{t}\|^2\right)$, where $\sigma$ was chosen by an automated procedure\cite{kernlab}. 

\section*{Funding}
This research was partially supported by project MTM2017-86061-C2-1-P (AEI/FEDER, UE), by grant PID2021-128314NB-I00 (MCIN/AEI/ 10.13039/501100011033/FEDER, UE), by the Ramon y Cajal Grant RYC2019-027534-I, by the Basque Government through the BERC 2018-2021 and the BMTF ‘‘Mathematical Modeling Applied to Health’’ Project, by the Spanish Ministry of Science and Innovation (MCIN) and AEI (BCAM Severo Ochoa accreditation CEX2021-001142-S), and by grants VA005P17 and VA002G18 (Junta de Castilla y León). The work of IB was financially supported in part by grants from the Departamento de Educaci\'on, Pol\'itica Ling\"u\'istica y Cultura del Gobierno Vasco [IT1456-22] and through the project [PID2020-115882RB-I00 / AEI / 10.13039/501100011033] funded by Agencia Estatal de Investigaci\'on and acronym ``S3M1P4R". This work was supported in part by the health outcomes group from Galdakao-Barrualde Health Organization; the Kronikgune Institute for Health Service Research and Biocruces Bizkaia Health Research Institute; and the thematic networks–REDISSEC (Red de Investigación en Servicios de Salud en Enfermedades Crónicas) and Network for Research on Chronicity, Primary Care, and Health Promotion (RICAPPS) – of the Instituto de Salud Carlos III.
\section*{Author contributions statement}
HI: conceptualisation, methodology, software, data curation, data analysis, visualization, original draft, review and editing. IBa and JMQ: conceptualisation, data acquisition, data curation, review and editing. PG: conceptualisation, review and editing. MXRA: conceptualisation, review and editing. IBe:  conceptualisation, review and editing. All authors approved the manuscript.

\section*{Data availability}
The datasets generated during and/or analysed during the current study are not publicly available due to the sensible nature of the information and the associated privacy issues but are available from the corresponding author on reasonable request.
The code for the methods presented in this work, alongside with meaningful synthetic examples, is freely available at 
\url{https://github.com/HristoInouzhe/Bias-from-Causality-perspective}.

\section*{Ethics declaration}
\noindent \textbf{Competing interests}. The authors declare no competing interests.




\bibliography{sample}

\section*{Appendix}
\begin{appendices}

\section{Unfairness due to nursing home status}\label{Ate_table}
\begin{table}[h!]
\centering
\begin{footnotesize}
\begin{tabular}{ccccccccc}
                    & \multicolumn{8}{c}{ATE: Hospitalisation}                                                                                  \\ \cline{2-9} 
                    & \multicolumn{2}{c}{Period 1} & \multicolumn{2}{c}{Period 2} & \multicolumn{2}{c}{Period 3} & \multicolumn{2}{c}{Period 4} \\ \cline{2-9} 
                    & Avg.      & CI    & Avg.      & CI    & Avg.      & CI    & Avg.      & CI    \\ \cline{2-9} 
Unmatched           & -0.37     & [-0.39,-0.35]             & -0.01     & [-0.03, 0.01]             & 0.00      & [-0.02, 0.02]             & 0.05      & [0.03, 0.07]             \\ \hline
Unmatched 2         & -0.354    & [-0.358, -0.35]            & -0.059    & [-0.06,-0.058]            & -0.069    & [-0.07, -0.068]            & 0.052     & [0.051, 0.053]            \\ \hline
Matched Euc         & -0.48     & [-0.50, -0.46]             & -0.15     & [-0.18, -0.12]             & -0.14     & [-0.17, -0.11]             & -0.03     & [-0.06, 0]             \\ \hline
Matched Euc 2       & -0.43     & [-0.47, -0.39]             & -0.14     & [-0.17, -0.11]             & -0.13     & [-0.16, -0.10]             & -0.04     & [-0.07, -0.01]             \\ \hline
Matched Prop        & -0.47     & [-0.49, -0.45]             & -0.15     & [-0.18, -0.12]             & -0.14     & [-0.17, -0.11]             & -0.05     & [-0.08, -0.02]             \\ \hline
Matched Prop 2      & -0.44     & [-0.47, -0.41]             & -0.15     & [-0.18, -0.12]             & -0.15     & [-0.18, -0.12]             & -0.05     & [-0.08, -0.02]             \\ \hline
Inverse Weighting & -0.31     & [-0.36, -0.26]             & -0.04     & [-0.09, 0.01]             & -0.04     & [-0.09, 0.01]             & 0.04      & [-0.01, 0.09]             \\ \hline
Inverse Weighting 2 & -0.36     & [-0.42, -0.30]             & -0.04     & [-0.09, 0.01]             & -0.06     & [-0.11, -0.01]             & 0.04      & [-0.01, 0.09]             \\ \hline
\end{tabular}
\vspace*{10pt}

\begin{tabular}{ccccccccc}
                    & \multicolumn{8}{c}{ATE: Mortality}                                                                                            \\ \cline{2-9} 
                    & \multicolumn{2}{c}{Period 1} & \multicolumn{2}{c}{Period 2} & \multicolumn{2}{c}{Period 3} & \multicolumn{2}{c}{Period 4} \\ \cline{2-9} 
                    & Avg.      & CI    & Avg.      & CI    & Avg.      & CI    & Avg.      & CI    \\ \cline{2-9} 
Unmatched           & 0.26      & [0.24, 0.28]             & 0.31      & [0.29, 0.33]             & 0.24      & [0.22, 0.26]             & 0.09      & [0.07, 0.11]             \\ \hline
Unmatched 2         & 0.079     & [0.078, 0.08]            & 0.147     & [0.146, 0.148]            & 0.097     & [0.096, 0.098]            & 0.037     & [0.036, 0.038]            \\ \hline
Matched Euc         & 0.18      & [0.16, 0.20]             & 0.13      & [0.10, 0.16]             & 0.07      & [0.04, 0.10]             & 0.03      & [0, 0.06]             \\ \hline
Matched Euc 2       & 0.09      & [0.05, 0.13]             & 0.14      & [0.11, 0.17]             & 0.07      & [0.04, 0.10]             & 0.02      & [-0.01, 0.05]             \\ \hline
Matched Prop        & 0.19      & [0.17, 0.21]             & 0.13      & [0.10, 0.16]             & 0.07      & [0.04, 0.10]             & 0.02      & [-0.01, 0.05]             \\ \hline
Matched Prop 2      & 0.09      & [0.06, 0.12]             & 0.12      & [0.09, 0.15]             & 0.05      & [0.02, 0.08]             & 0.02      & [-0.01, 0.05]             \\ \hline
Inverse Weighting & 0.10      & [0.05, 0.15]             & 0.14      & [0.10, 0.18]             & 0.12      & [0.07, 0.17]             & 0.03      & [-0.01, 0.07]             \\ \hline
Inverse Weighting 2 & 0.09      & [0.02, 0.16]             & 0.15      & [0.11, 0.19]             & 0.11      & [0.05, 0.17]             & 0.04      & [0, 0.08]             \\ \hline
\end{tabular}
\vspace*{10pt}

\begin{tabular}{ccccccccc}
                    & \multicolumn{8}{c}{ATE: In-hospital mortality}                                                                                \\ \cline{2-9} 
                    & \multicolumn{2}{c}{Period 1} & \multicolumn{2}{c}{Period 2} & \multicolumn{2}{c}{Period 3} & \multicolumn{2}{c}{Period 4} \\ \cline{2-9} 
                    & Avg.      & CI    & Avg.      & CI    & Avg.      & CI    & Avg.      & CI    \\ \cline{2-9} 
Unmatched           & 0.40      & [0.36, 0.44]             & 0.42      & [0.38, 0.46]             & 0.31      & [0.25, 0.37]             & 0.15      & [0.05, 0.25]             \\ \hline
Unmatched 2         & 0.209     & [0.204, 0.214]            & 0.248     & [0.243, 0.253]            & 0.127     & [0.124, 0.13]            & 0.066     & [0.062, 0.07]            \\ \hline
Matched Euc         & 0.28      & [0.23, 0.33]             & 0.28      & [0.22, 0.34]             & 0.14      & [0.06, 0.22]             & 0.13      & [0, 0.26]             \\ \hline
Matched Euc 2       & 0.30      & [0.24, 0.36]             & 0.30      & [0.23, 0.37]             & 0.12      & [0.04, 0.20]             & 0.16      & [0.02, 0.30]             \\ \hline
Matched Prop        & 0.21      & [0.16, 0.26]             & 0.26      & [0.20, 0.32]             & 0.12      & [0.04, 0.20]             & 0.16      & [0.03, 0.29]             \\ \hline
Matched Prop 2      & 0.23      & [0.17, 0.29]             & 0.27      & [0.21, 0.33]             & 0.10      & [0.02, 0.18]             & 0.18      & [0.05, 0.31]             \\ \hline
Inverse Weighting & 0.29      & [0.17, 0.41]             & 0.23      & [0.13, 0.33]             & 0.19      & [0.02, 0.36]             & 0.15      & [-0.12, 0.42]             \\ \hline
Inverse Weighting 2 & 0.27      & [0.08, 0.48]             & 0.32      & [0.20, 0.44]             & 0.27      & [0.01, 0.53]             & 0.16      & [-0.11, 0.43]             \\ \hline
\end{tabular}
\end{footnotesize}
\caption{ATE point estimates for four different periods, eight different estimators, and three different outcomes: hospitalisation, mortality, and in-hospital mortality. Estimators' point estimates are presented in Avg., while the 95\% confidence interval is given in CI.}
\end{table}
\clearpage
\section{Quality of ATE estimation}\label{Quality_ATE_estiamtion}
\begin{figure*}[h!]
    \centering
    \includegraphics[scale=0.4]{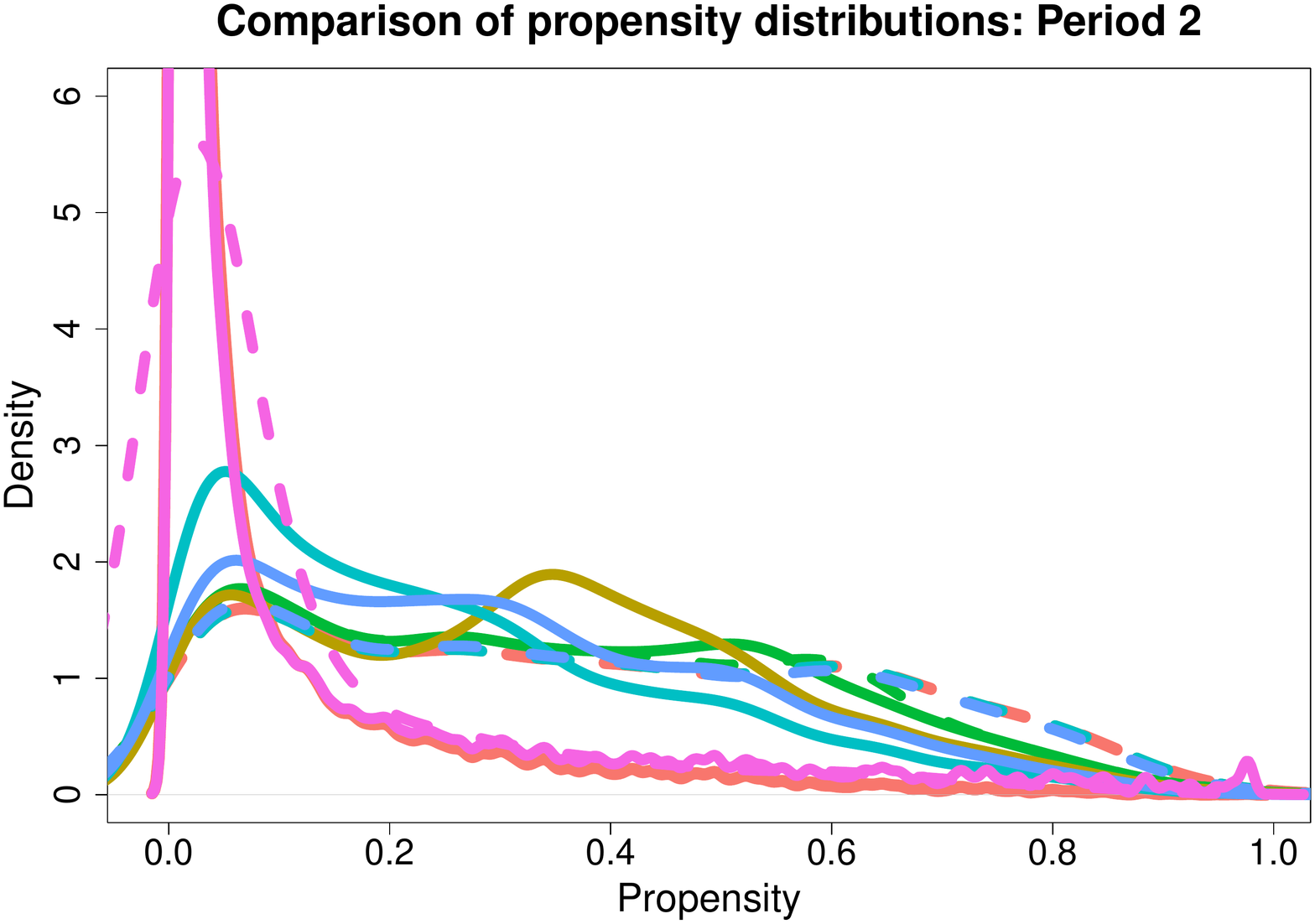}
    \includegraphics[scale=0.4]{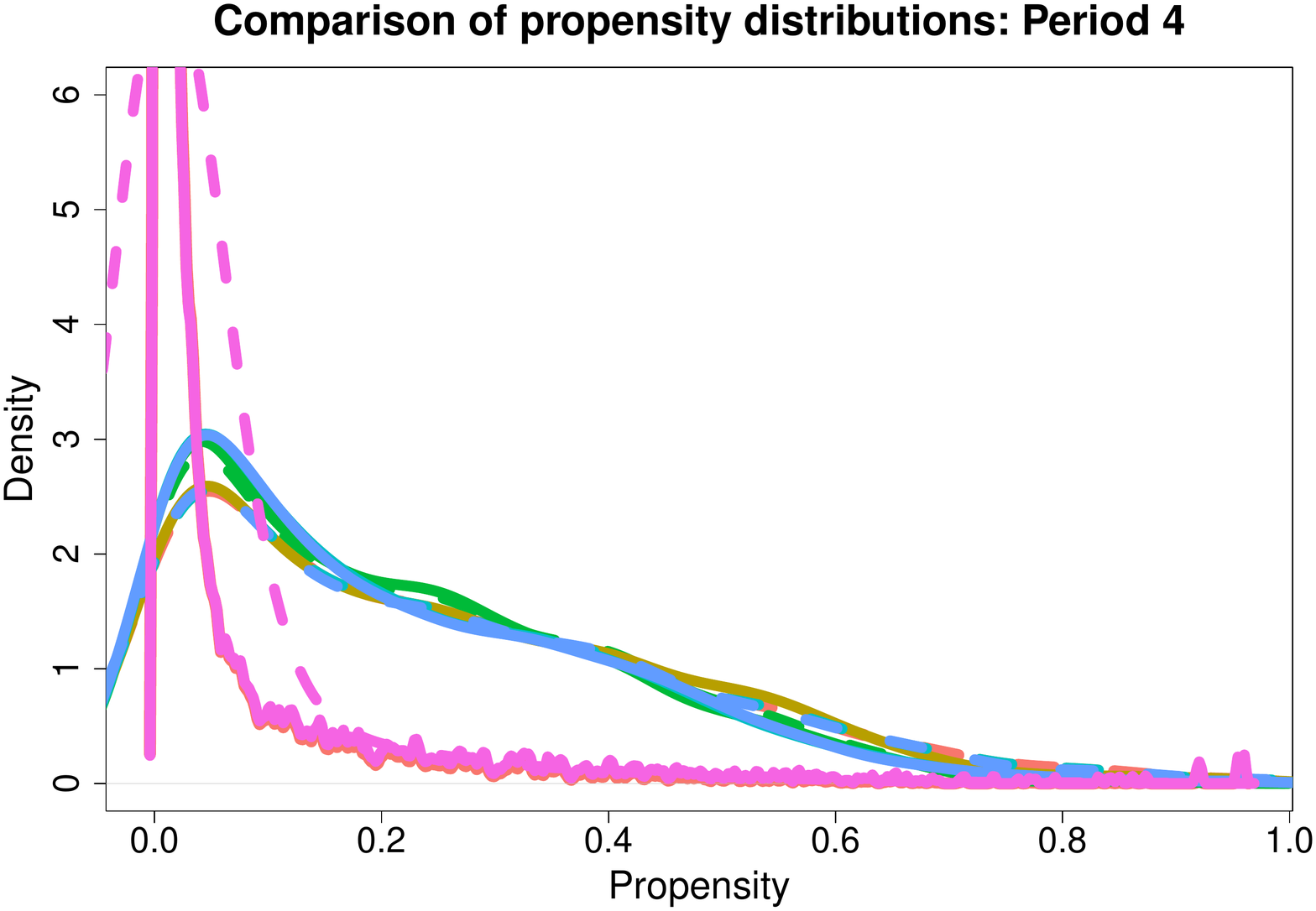}
    \includegraphics[scale = 0.4]{Full_comparison_prop_legend.png}
    \caption{Distribution of propensity score estimates for \emph{outcomes hospitalisation and mortality} for nursing home residents, $Z=1$, in dashed, and non-residents, $Z=0$, in solid. Methods correspond with Table \ref{table:characterization_estimations} in the main document, where both unmatched methods produce the same propensity scores, which is also the case for the inverse weightings.}
    \label{fig_supplementary:prop_comparison}
\end{figure*}
\begin{figure*}[h!]
    \centering
    \includegraphics[scale=0.27]{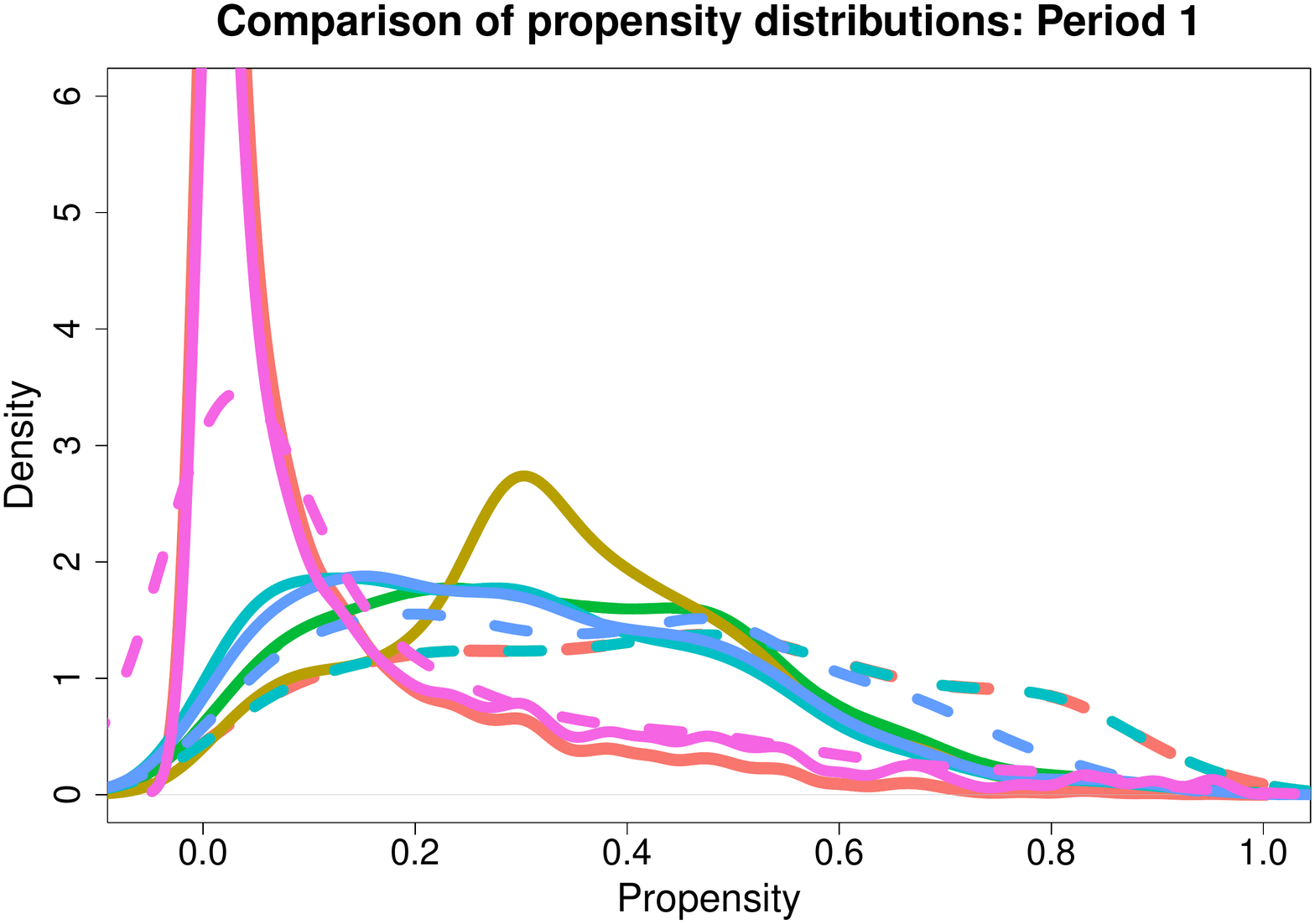}\includegraphics[scale=0.27]{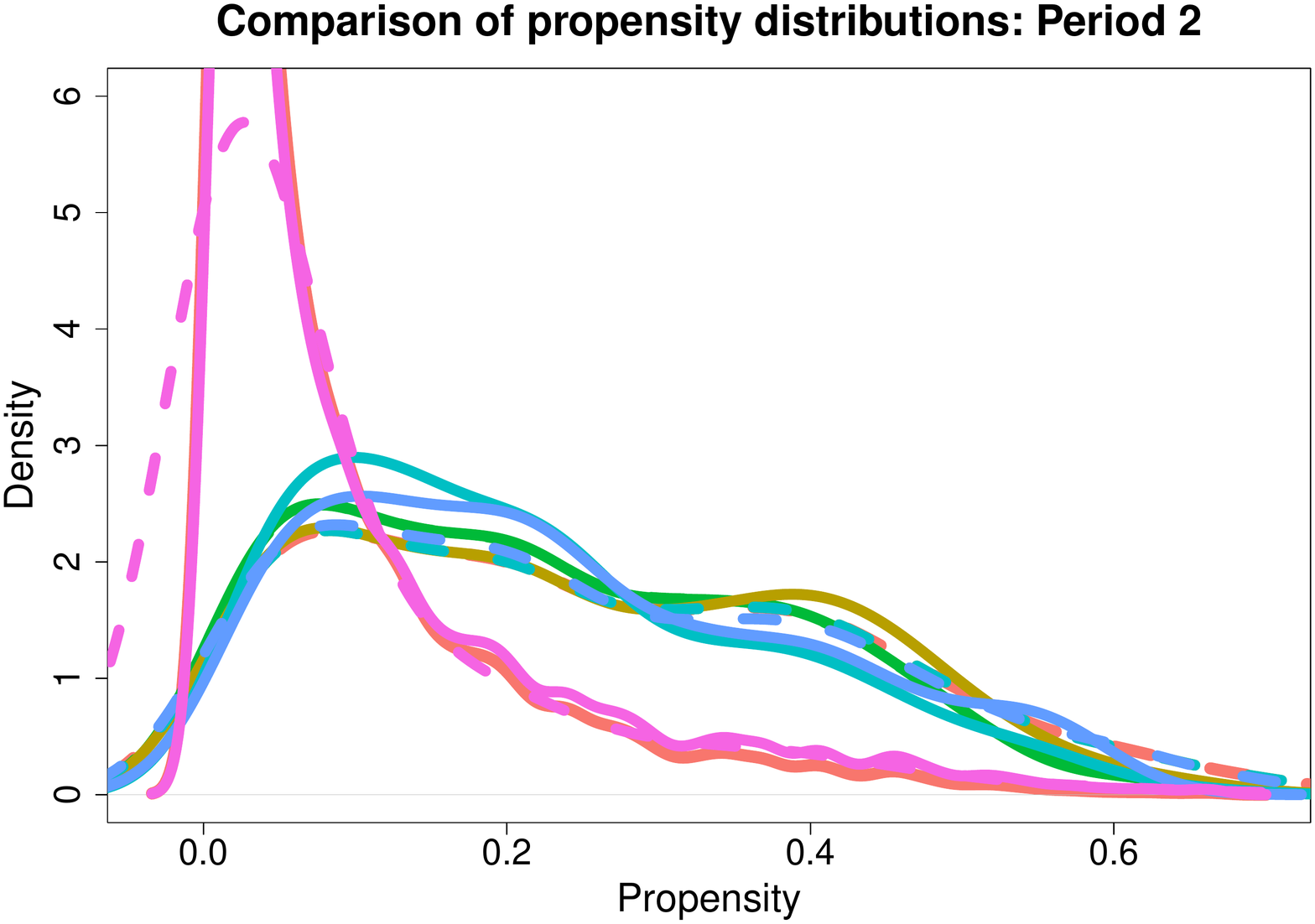}
    \includegraphics[scale=0.27]{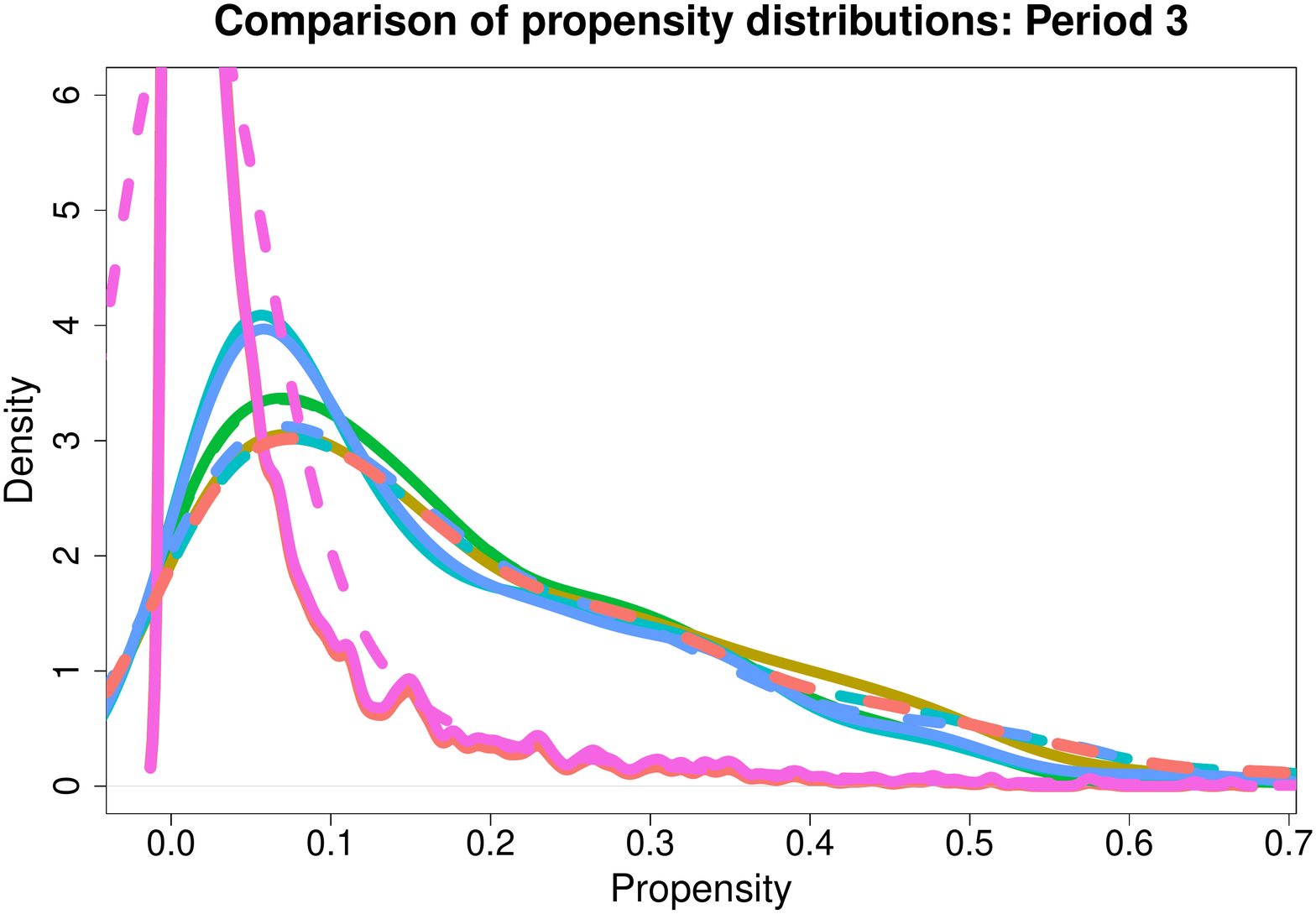}\includegraphics[scale=0.27]{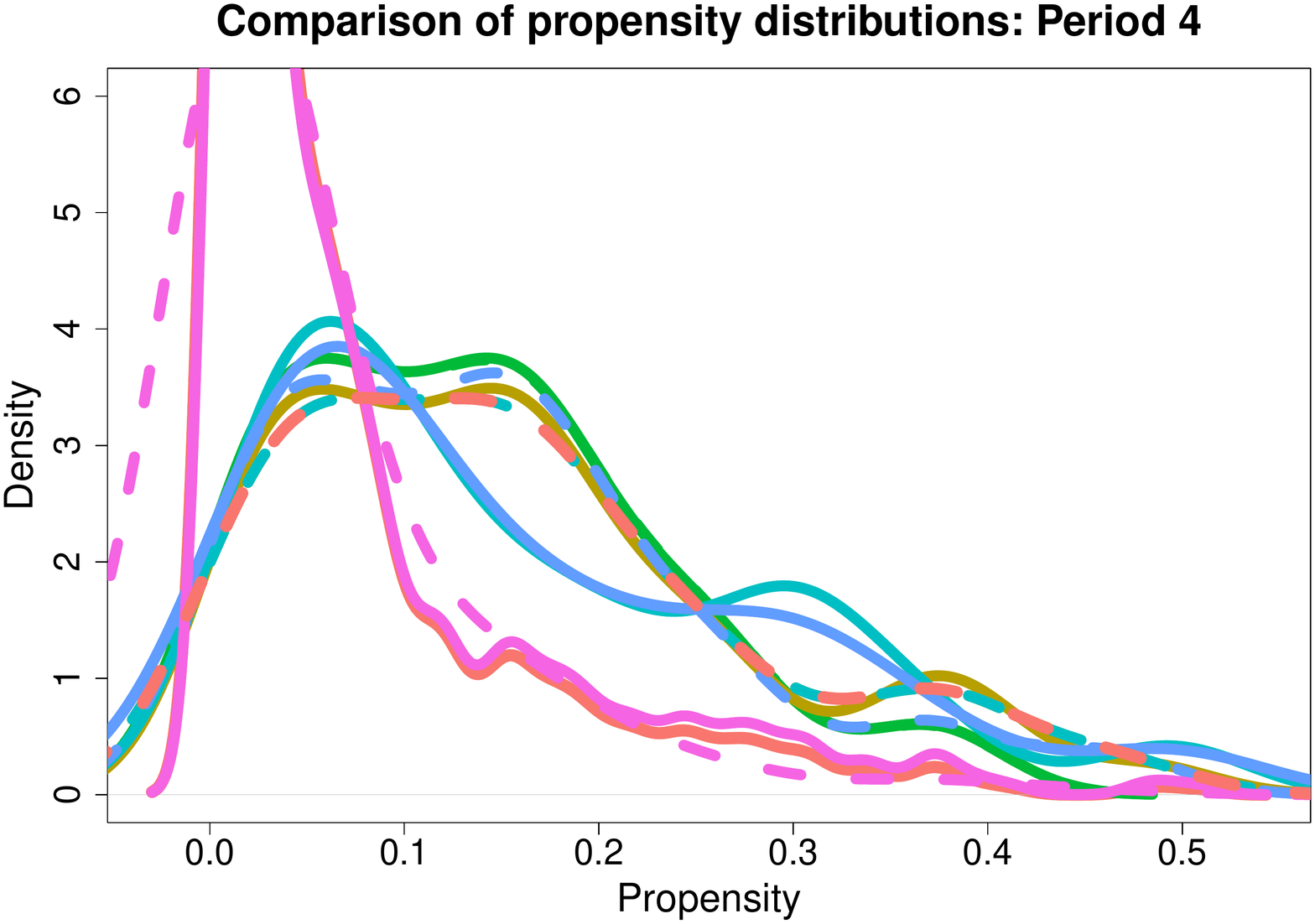}
    \includegraphics[scale = 0.4]{Full_comparison_prop_legend.png}
    \caption{Distribution of propensity score estimates for the data used in \emph{outcome in-hospital mortality} for nursing home residents, $Z=1$, in dashed, and non-residents, $Z=0$, in solid. Methods correspond with Table \ref{table:characterization_estimations} in the main document, where both unmatched methods produce the same propensity scores, which is also the case for the inverse weightings.}
    \label{fig_supplementary:prop_comparison_hosp}
\end{figure*}
\begin{figure*}[h!]
    \centering
    \includegraphics[scale=0.37]{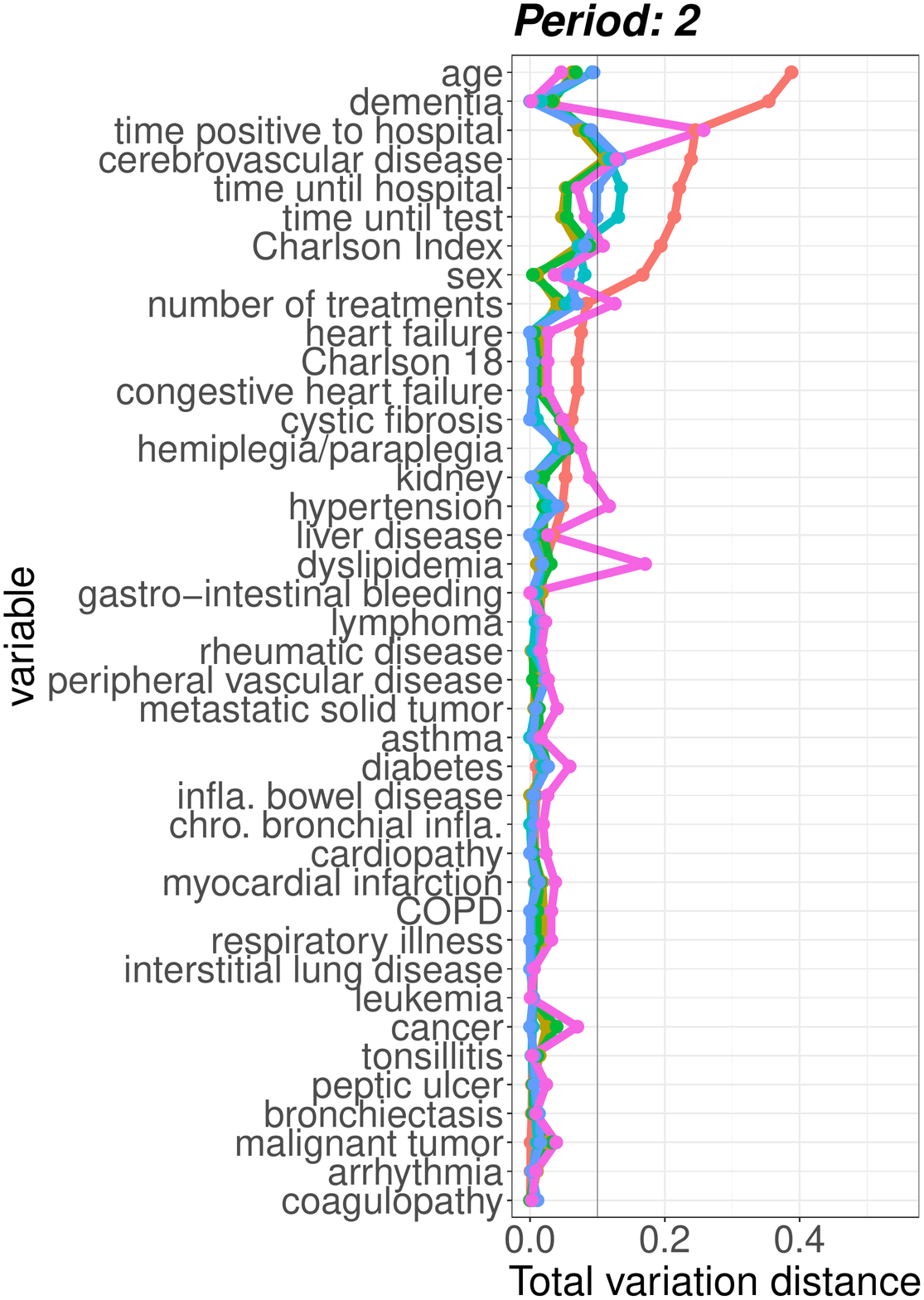}\includegraphics[scale=0.37]{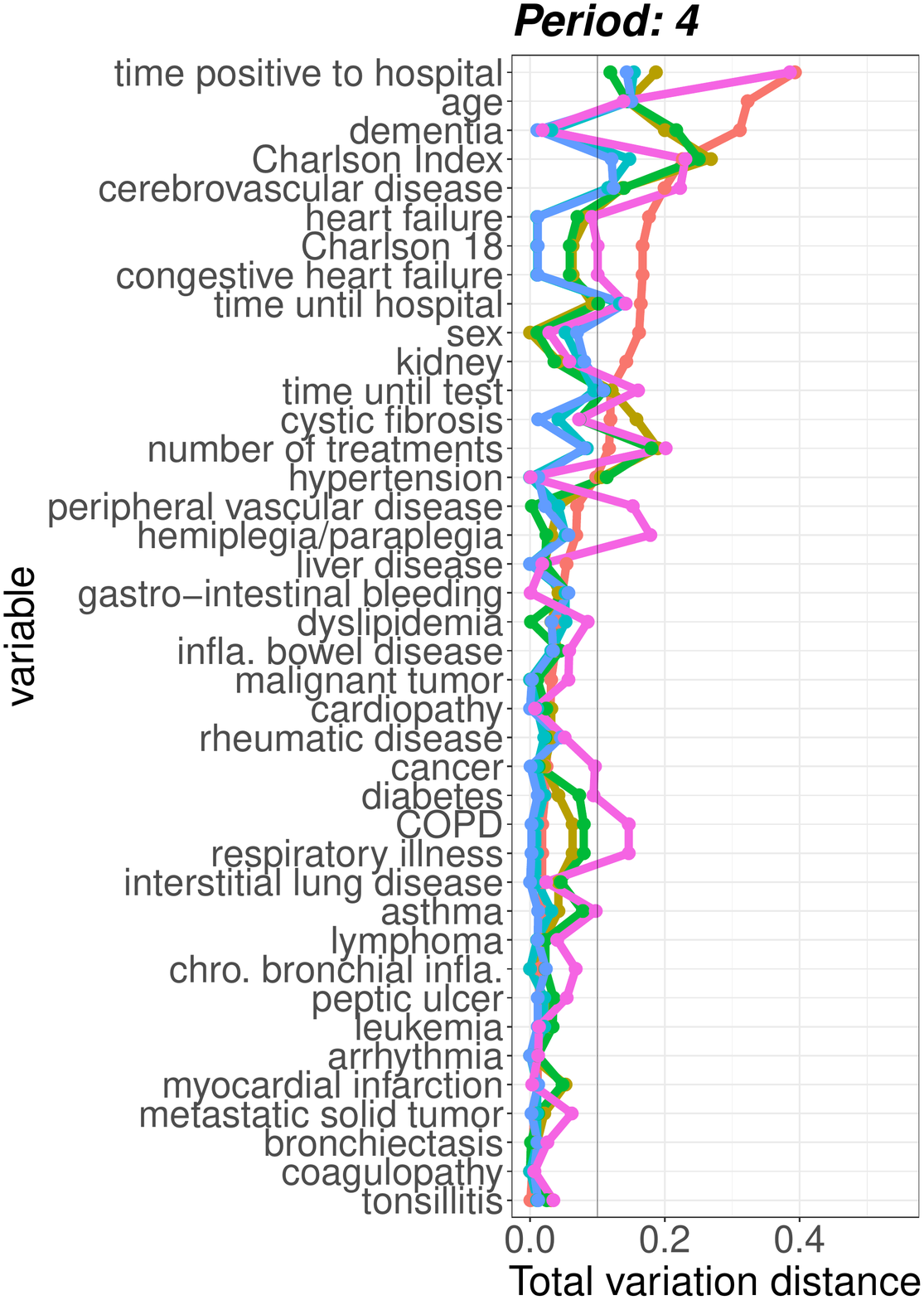}
    \includegraphics[scale = 0.35]{Full_comparison_prop_legend_2.png}
    \caption{Estimates of the Total Variation distance between the marginals corresponding to nursing home residents and non-residents for \emph{outcomes hospitalisation and mortality}.}
    \label{fig_supplementary:dtv_comparison}
\end{figure*}
\begin{figure*}[h!]
    \centering
    \includegraphics[scale=0.34]{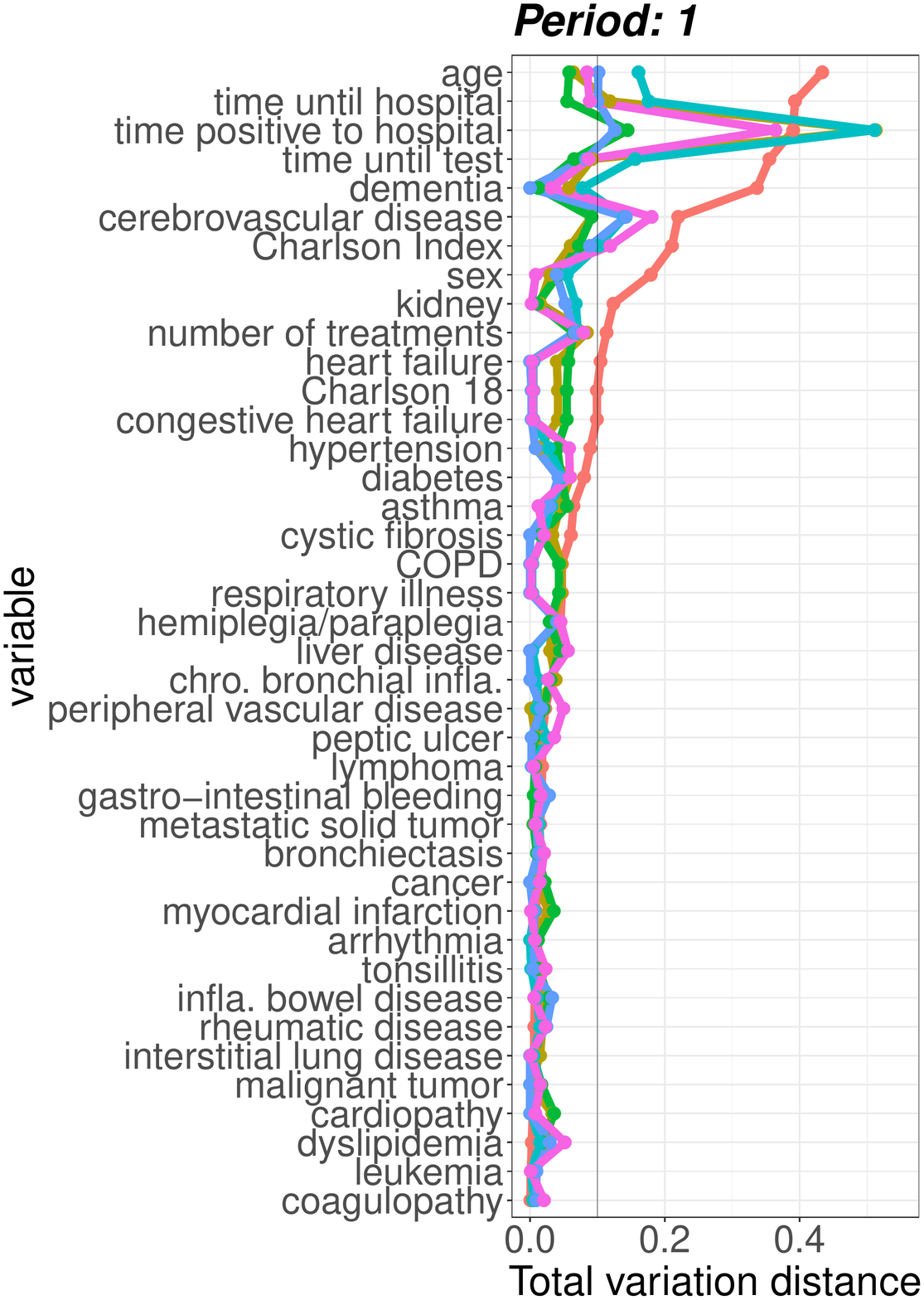}\includegraphics[scale=0.34]{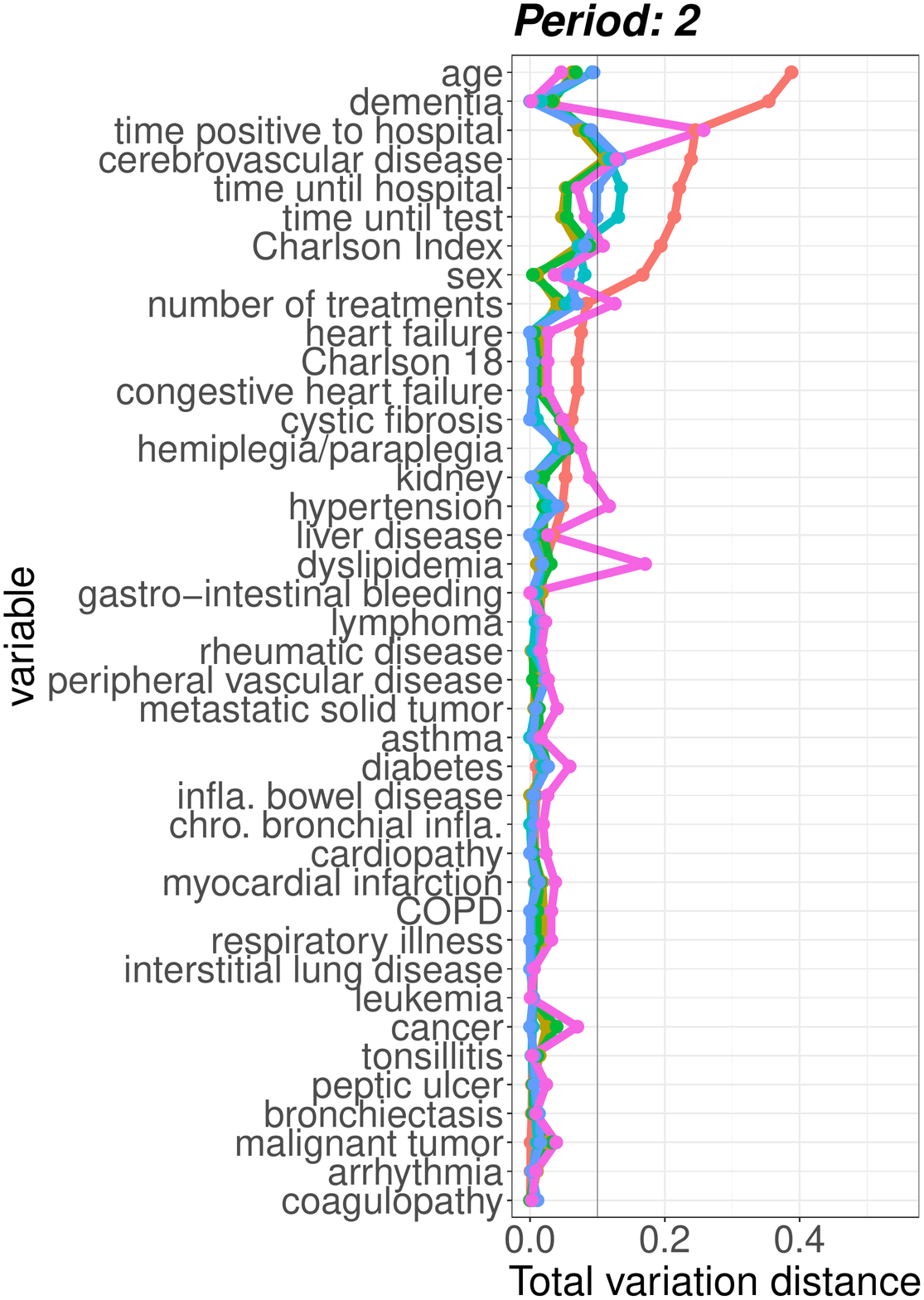}
    \includegraphics[scale=0.34]{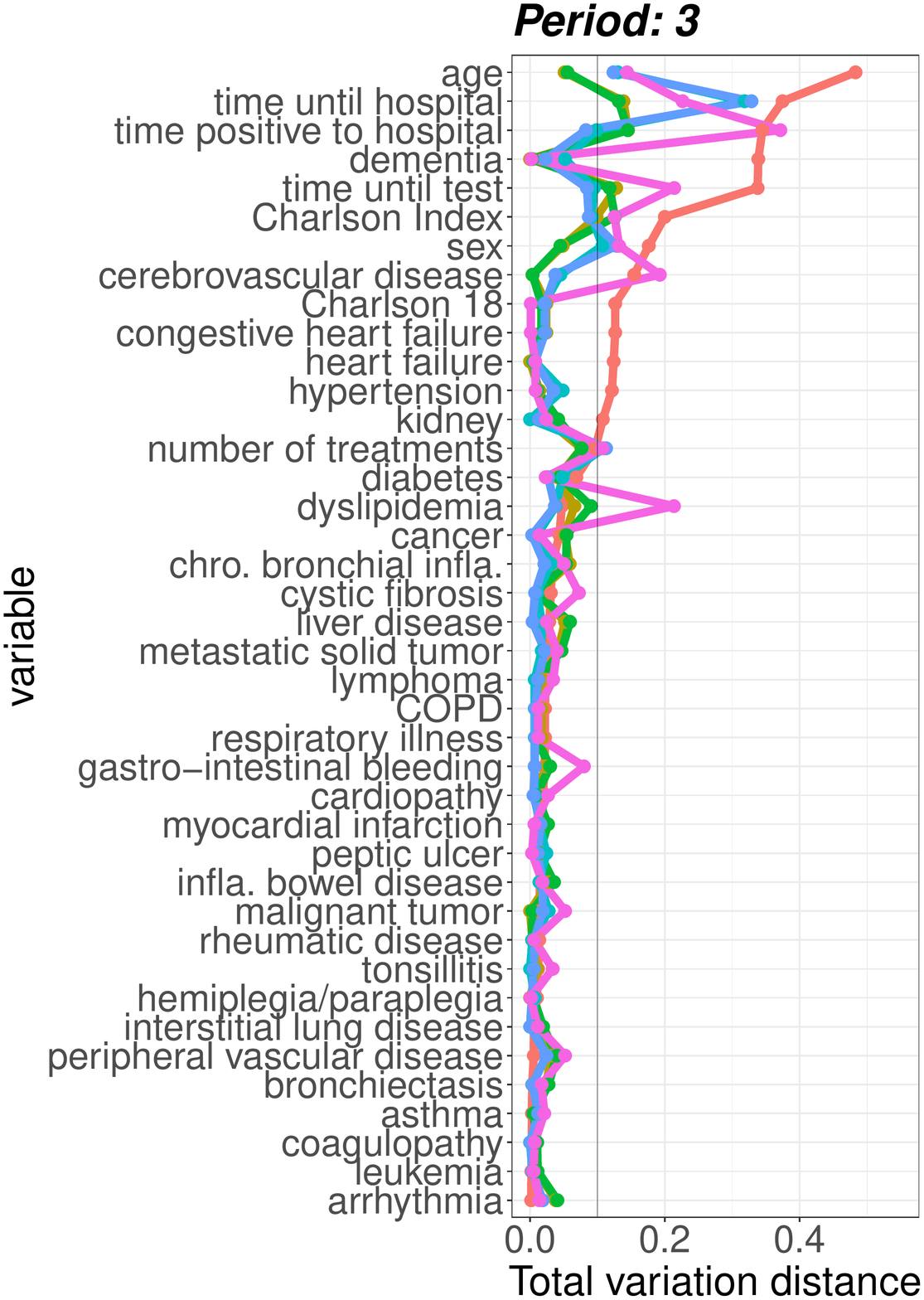}\includegraphics[scale=0.34]{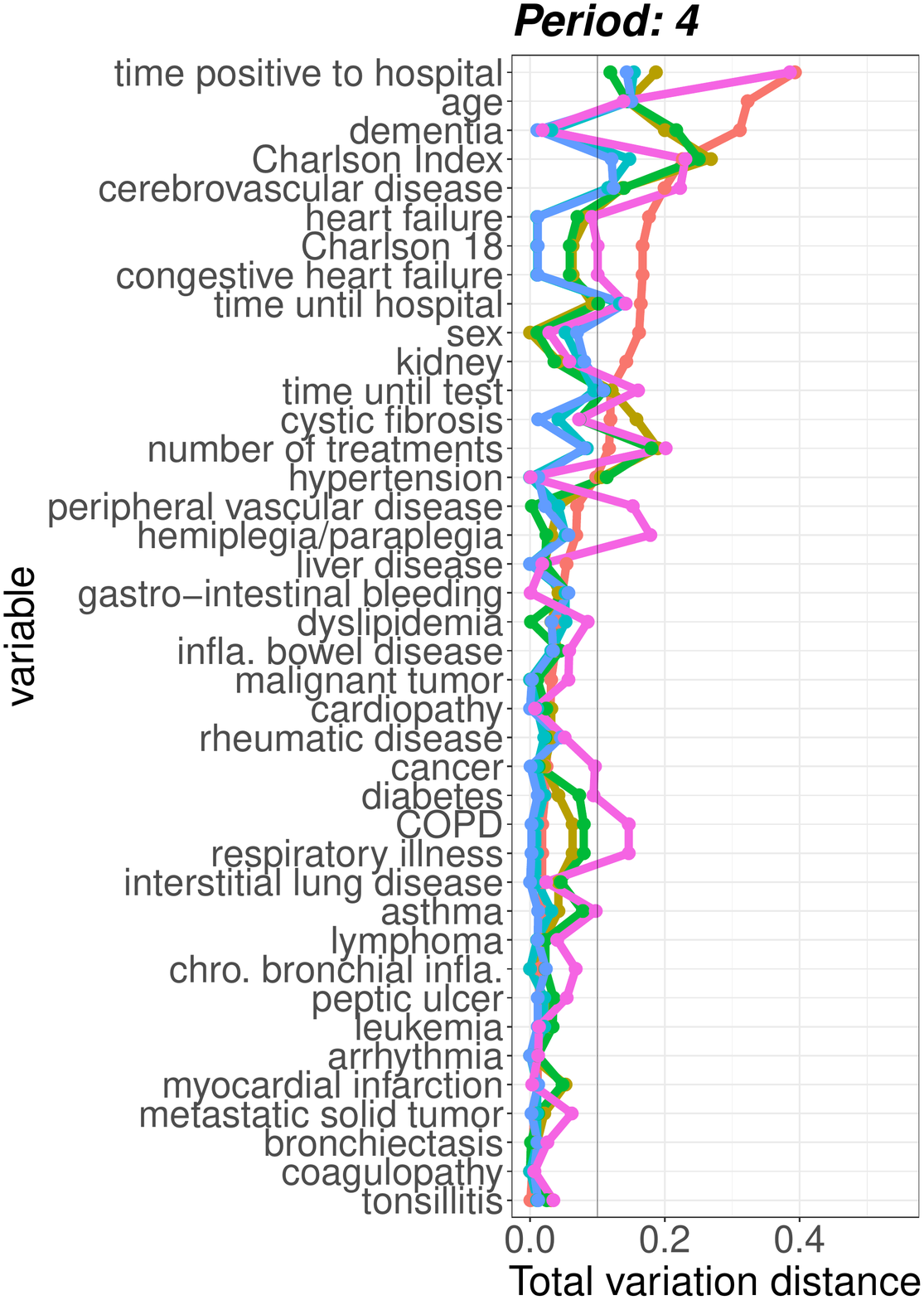}
    \includegraphics[scale = 0.35]{Full_comparison_prop_legend_2.png}
    \caption{Estimates of the Total Variation distance between the marginals corresponding to nursing home residents and non-residents for \emph{outcome in-hospital mortality}.}
    \label{fig_supplementary:dtv_comparison_hosp}
\end{figure*}

\begin{table}[h!]
\begin{scriptsize}
\begin{center}
    \begin{tabular}{ccccccccc}
               & \multicolumn{8}{c}{Reject $H_0$: Hospitalisation, Mortality}                                                                  \\ \cline{2-9} 
               & \multicolumn{2}{c}{Period 1} & \multicolumn{2}{c}{Period 2} & \multicolumn{2}{c}{Period 3} & \multicolumn{2}{c}{Period 4} \\ \cline{2-9} 
               & MMD      & MMD Asym   & MMD      & MMD Asym     & MMD      & MMD Asym    & MMD      & MMD Asym  \\ \cline{2-9} 
Unmatched      & TRUE     & TRUE              & TRUE     & NONE              & TRUE     & NONE              & TRUE     & NONE              \\ \hline
Matched Euc    & TRUE     & TRUE              & FALSE    & TRUE              & FALSE    & TRUE              & FALSE    & TRUE              \\ \hline
Matched Euc 2  & FALSE    & TRUE              & FALSE    & TRUE              & FALSE    & TRUE              & FALSE    & FALSE             \\ \hline
Matched Prop   & TRUE     & TRUE              & FALSE    & TRUE              & FALSE    & TRUE              & FALSE    & FALSE             \\ \hline
Matched Prop 2 & FALSE    & TRUE              & FALSE    & FALSE             & FALSE    & FALSE             & FALSE    & TRUE              \\ \hline
\end{tabular}
\vspace*{10pt}

\begin{tabular}{ccccccccc}
               & \multicolumn{8}{c}{Reject $H_0$: In-hospital mortality}                                                                       \\ \cline{2-9} 
               & \multicolumn{2}{c}{Period 1} & \multicolumn{2}{c}{Period 2} & \multicolumn{2}{c}{Period 3} & \multicolumn{2}{c}{Period 4} \\ \cline{2-9} 
               & MMD      & MMD Asym    & MMD      & MMD Asym    & MMD      & MMD Asym    & MMD      & MMD Asym    \\ \cline{2-9} 
Unmatched      & FALSE    & TRUE              & FALSE    & TRUE              & FALSE    & TRUE              & FALSE    & TRUE              \\ \hline
Matched Euc    & FALSE    & TRUE              & FALSE    & FALSE             & FALSE    & FALSE             & FALSE    & FALSE             \\ \hline
Matched Euc 2  & FALSE    & FALSE             & FALSE    & FALSE             & FALSE    & FALSE             & FALSE    & FALSE             \\ \hline
Matched Prop   & FALSE    & TRUE              & FALSE    & FALSE             & FALSE    & FALSE             & FALSE    & FALSE             \\ \hline
Matched Prop 2 & FALSE    & TRUE              & FALSE    & FALSE             & FALSE    & FALSE             & FALSE    & FALSE             \\ \hline
\end{tabular}
\end{center}
\end{scriptsize}
\caption{Results of the Maximum Mean Discrepancy goodness of fit test for two different statistics, MMD as in Table \ref{table:MMD_h0_rejection} in the main document and an alternative computation based on asymptotics and bootstrapping \cite{gretton2006kernel,kernlab}. Entries NONE denote that we did not compute the statistic due to relatively long running time ($>$1h).}
\label{tab_supplementary:H0_rejection}
\end{table}

\begin{table*}[h!]
\centering
\begin{tabular}{ccccc}
                          & \multicolumn{4}{c}{Distance weights $\mathbf{w}$: In-hospital mortaility} \\ \cline{2-5} 
                          & Period 1        & Period 2        & Period 3        & Period 4       \\ \cline{2-5} 
age                       & 1               & 1               & 2               & 1              \\ \hline
Charlson index            & 2               & 1               & 1               & 3              \\ \hline
time until test           & 2               & 1               & 1               & 2              \\ \hline
time until hospital       & 2               & 1               & 1               & 1              \\ \hline
time positive to hospital & 5               & 4               & 8               & 1              \\ \hline
cancer                    & 1               & 1               & 8               & 6              \\ \hline
respiratory illness       & 1               & 1               & 1               & 1              \\ \hline
cardiopathy               & 1               & 1               & 1               & 1              \\ \hline
heart failure             & 1               & 1               & 1               & 1              \\ \hline
interstitial lung disease & 1               & 1               & 1               & 1              \\ \hline
liver disease             & 1               & 1               & 1               & 1              \\ \hline
cystic fibrosis           & 1               & 1               & 1               & 1              \\ \hline
dementia                  & 1               & 1               & 1               & 1              \\ \hline
\end{tabular}
\caption{Weights, $\mathbf{w}$, used to compute the weighted Euclidean distance for matching. For the data employed for outputs hospitalisation and mortality, it was all ones, i.e., the standard Euclidean distance. For the outcome in-hospital mortality, we performed several trials to improve the resulting total variation distance in the one-dimensional marginals, and picked the best performance. The procedure can be much improved and formalised, but this was out of the scope of this work.}
\label{tab_supplementary:weights}
\end{table*}
\clearpage
\section{Descriptive analysis}\label{descriptive_analysis}
\newgeometry{left=0.1cm,bottom=1.5cm}
\begin{landscape}
\begin{table}[]
\begin{scriptsize}
\begin{tabular}{|c|clccc|cccc|}
\hline
                                                                                            & \multicolumn{5}{c|}{\textbf{Patients at nursing homes (\textgreater{}= 60 years old)}}                                                                                 & \multicolumn{4}{c|}{\textbf{Not nursing homes (\textgreater{}= 60 years old)}}                                                                                             \\ \hline
\textbf{Variables}                                                                          & \multicolumn{2}{c|}{\textbf{Period 1}}        & \multicolumn{1}{c|}{\textbf{Period 2}}        & \multicolumn{1}{c|}{\textbf{Period 3}}        & \textbf{Omicron}       & \multicolumn{1}{c|}{\textbf{Period 1}}       & \multicolumn{1}{c|}{\textbf{Period 2}}         & \multicolumn{1}{c|}{\textbf{Period 3}}         & \textbf{Omicron}          \\ \hline
\textbf{TOTAL}                                                                              & \multicolumn{2}{c|}{\textbf{3,370 (41.96\%)}} & \multicolumn{1}{c|}{\textbf{2,314 (28.81\%)}} & \multicolumn{1}{c|}{\textbf{1,429 (17.79\%)}} & \textbf{919 (11.44\%)} & \multicolumn{1}{c|}{\textbf{5,845 (7.51\%)}} & \multicolumn{1}{c|}{\textbf{19,145 (24.61\%)}} & \multicolumn{1}{c|}{\textbf{30,383 (39.06\%)}} & \textbf{22,406 (28.81\%)} \\ \hline
\textit{Sociodemographic variables}                                                         & \multicolumn{2}{c|}{}                         & \multicolumn{1}{c|}{}                         & \multicolumn{1}{c|}{}                         &                        & \multicolumn{1}{c|}{}                        & \multicolumn{1}{c|}{}                          & \multicolumn{1}{c|}{}                          &                           \\ \hline
\textbf{Gender, N (\%)$^\mathrm{All}$}                                                      & \multicolumn{2}{c|}{}                         & \multicolumn{1}{c|}{}                         & \multicolumn{1}{c|}{}                         &                        & \multicolumn{1}{c|}{}                        & \multicolumn{1}{c|}{}                          & \multicolumn{1}{c|}{}                          &                           \\ \hline
Female                                                                                      & \multicolumn{2}{c|}{2,370 (70.33)}            & \multicolumn{1}{c|}{1,613 (69.71)}            & \multicolumn{1}{c|}{997 (69.77)}              & 630 (68.55)            & \multicolumn{1}{c|}{2,982 (51.02)}           & \multicolumn{1}{c|}{10,168 (53.11)}            & \multicolumn{1}{c|}{15,956 (52.52)}            & 12,104 (54.02)            \\ \hline
Male                                                                                        & \multicolumn{2}{c|}{1,000 (29.67)}            & \multicolumn{1}{c|}{701 (30.29)}              & \multicolumn{1}{c|}{432 (30.23)}              & 289 (31.45)            & \multicolumn{1}{c|}{2,863 (48.98)}           & \multicolumn{1}{c|}{8,977 (46.89)}             & \multicolumn{1}{c|}{14,427 (47.48)}            & 10,302 (45.98)            \\ \hline
\textbf{Age, Median {[}Q1,Q3{]}$^\mathrm{All}$}                                             & \multicolumn{2}{c|}{87 {[}81,91{]}}           & \multicolumn{1}{c|}{87 {[}80,91{]}}           & \multicolumn{1}{c|}{87 {[}81,91{]}}           & 88 {[}81,92{]}         & \multicolumn{1}{c|}{71 {[}63,80{]}}          & \multicolumn{1}{c|}{70 {[}64,80{]}}            & \multicolumn{1}{c|}{69 {[}64,77{]}}            & 69 {[}63,76{]}            \\ \hline
\textbf{Vaccines (2-3 doses), N (\%)$^\mathrm{All}$}                                        & \multicolumn{2}{c|}{}                         & \multicolumn{1}{c|}{}                         & \multicolumn{1}{c|}{}                         &                        & \multicolumn{1}{c|}{}                        & \multicolumn{1}{c|}{}                          & \multicolumn{1}{c|}{}                          &                           \\ \hline
0 doses                                                                                     & \multicolumn{2}{c|}{3,370 (100.00)}           & \multicolumn{1}{c|}{2,314 (100.00)}           & \multicolumn{1}{c|}{571 (39.96)}              & 17 (1.85)              & \multicolumn{1}{c|}{5,845 (100.00)}          & \multicolumn{1}{c|}{19,145 (100.00)}           & \multicolumn{1}{c|}{16,220 (53.39)}            & 1,299 (5.80)              \\ \hline
1 dose                                                                                      & \multicolumn{2}{c|}{0 (0.00)}                 & \multicolumn{1}{c|}{0 (0.00)}                 & \multicolumn{1}{c|}{143 (10.01)}              & 10 (1.09)              & \multicolumn{1}{c|}{0 (0.00)}                & \multicolumn{1}{c|}{0 (0.00)}                  & \multicolumn{1}{c|}{2,148 (7.07)}              & 258 (1.15)                \\ \hline
2 doses                                                                                     & \multicolumn{2}{c|}{0 (0.00)}                 & \multicolumn{1}{c|}{0 (0.00)}                 & \multicolumn{1}{c|}{545 (38.14)}              & 60 (6.53)              & \multicolumn{1}{c|}{0 (0.00)}                & \multicolumn{1}{c|}{0 (0.00)}                  & \multicolumn{1}{c|}{10,952 (36.05)}            & 5,821 (25.98)             \\ \hline
3 doses                                                                                     & \multicolumn{2}{c|}{0 (0.00)}                 & \multicolumn{1}{c|}{0 (0.00)}                 & \multicolumn{1}{c|}{170 (11.90)}              & 832 (90.53)            & \multicolumn{1}{c|}{0 (0.00)}                & \multicolumn{1}{c|}{0 (0.00)}                  & \multicolumn{1}{c|}{1,063 (3.50)}              & 15,028 (67.07)            \\ \hline
\textit{Comorbidities}                                                                      & \multicolumn{2}{c|}{}                         & \multicolumn{1}{c|}{}                         & \multicolumn{1}{c|}{}                         &                        & \multicolumn{1}{c|}{}                        & \multicolumn{1}{c|}{}                          & \multicolumn{1}{c|}{}                          &                           \\ \hline
\textbf{Charlson index, Median {[}Q1,Q3{]}$^\mathrm{All}$}                                  & \multicolumn{2}{c|}{2 {[}1,4{]}}              & \multicolumn{1}{c|}{2 {[}1,4{]}}              & \multicolumn{1}{c|}{2 {[}1,4{]}}              & 2 {[}1,4{]}            & \multicolumn{1}{c|}{1 {[}0,3{]}}             & \multicolumn{1}{c|}{1 {[}0,2{]}}               & \multicolumn{1}{c|}{1 {[}0,2{]}}               & 1 {[}0,2{]}               \\ \hline
\textbf{\begin{tabular}[c]{@{}c@{}}Myocardial infarction\\ , N (\%)$^{1,2,3}$\end{tabular}} & \multicolumn{2}{c|}{312 (9.26)}               & \multicolumn{1}{c|}{180 (7.78)}               & \multicolumn{1}{c|}{122 (8.54)}               & 83 (9.03)              & \multicolumn{1}{c|}{514 (8.79)}              & \multicolumn{1}{c|}{1,162 (6.07)}              & \multicolumn{1}{c|}{1,656 (5.45)}              & 1,204 (5.37)              \\ \hline
\textbf{Congestive heart failure, N (\%)$^\mathrm{All}$}                                    & \multicolumn{2}{c|}{791 (23.47)}              & \multicolumn{1}{c|}{499 (21.56)}              & \multicolumn{1}{c|}{321 (22.46)}              & 194 (21.11)            & \multicolumn{1}{c|}{814 (13.93)}             & \multicolumn{1}{c|}{1,871 (9.77)}              & \multicolumn{1}{c|}{2,579 (8.49)}              & 1,563 (6.98)              \\ \hline
\textbf{Peripheral vascular disease, N (\%)$^{1,3,4}$}                                      & \multicolumn{2}{c|}{382 (11.34)}              & \multicolumn{1}{c|}{281 (12.14)}              & \multicolumn{1}{c|}{133 (9.31)}               & 114 (12.40)            & \multicolumn{1}{c|}{586 (10.03)}             & \multicolumn{1}{c|}{1,598 (8.35)}              & \multicolumn{1}{c|}{2,207 (7.26)}              & 1,513 (6.75)              \\ \hline
\textbf{Cerebrovascular disease, N (\%)$^\mathrm{All}$}                                     & \multicolumn{2}{c|}{1,221 (36.23)}            & \multicolumn{1}{c|}{831 (35.91)}              & \multicolumn{1}{c|}{467 (32.68)}              & 353 (38.41)            & \multicolumn{1}{c|}{919 (15.72)}             & \multicolumn{1}{c|}{2,710 (14.16)}             & \multicolumn{1}{c|}{4,027 (13.25)}             & 2,781 (12.41)             \\ \hline
\textbf{Dementia, N (\%)$^\mathrm{All}$}                                                    & \multicolumn{2}{c|}{1,331 (39.50)}            & \multicolumn{1}{c|}{960 (41.49)}              & \multicolumn{1}{c|}{561 (39.26)}              & 354 (38.52)            & \multicolumn{1}{c|}{332 (5.68)}              & \multicolumn{1}{c|}{1,045 (5.46)}              & \multicolumn{1}{c|}{1,181 (3.89)}              & 720 (3.21)                \\ \hline
\textbf{COPD, N (\%)$^{1}$}                                                                 & \multicolumn{2}{c|}{743 (22.05)}              & \multicolumn{1}{c|}{540 (23.34)}              & \multicolumn{1}{c|}{331 (23.16)}              & 211 (22.96)            & \multicolumn{1}{c|}{1,412 (24.16)}           & \multicolumn{1}{c|}{3,791 (19.80)}             & \multicolumn{1}{c|}{5,565 (18.32)}             & 3,930 (17.54)             \\ \hline
\textbf{Rheumatic disease, N (\%)}                                                          & \multicolumn{2}{c|}{185 (5.49)}               & \multicolumn{1}{c|}{130 (5.62)}               & \multicolumn{1}{c|}{77 (5.39)}                & 42 (4.57)              & \multicolumn{1}{c|}{307 (5.25)}              & \multicolumn{1}{c|}{829 (4.33)}                & \multicolumn{1}{c|}{1,253 (4.12)}              & 994 (4.44)                \\ \hline
\textbf{Peptic ulcer, N (\%)$^{2}$}                                                         & \multicolumn{2}{c|}{223 (6.62)}               & \multicolumn{1}{c|}{148 (6.40)}               & \multicolumn{1}{c|}{88 (6.16)}                & 57 (6.20)              & \multicolumn{1}{c|}{348 (5.95)}              & \multicolumn{1}{c|}{925 (4.83)}                & \multicolumn{1}{c|}{1,455 (4.79)}              & 1,006 (4.49)              \\ \hline
\textbf{Liver disease, N (\%)}                                                              & \multicolumn{2}{c|}{}                         & \multicolumn{1}{c|}{}                         & \multicolumn{1}{c|}{}                         &                        & \multicolumn{1}{c|}{}                        & \multicolumn{1}{c|}{}                          & \multicolumn{1}{c|}{}                          &                           \\ \hline
Mild                                                                                        & \multicolumn{2}{c|}{246 (7.30)}               & \multicolumn{1}{c|}{146 (6.31)}               & \multicolumn{1}{c|}{110 (7.70)}               & 61 (6.64)              & \multicolumn{1}{c|}{486 (8.31)}              & \multicolumn{1}{c|}{1,338 (6.99)}              & \multicolumn{1}{c|}{2,102 (6.92)}              & 1,560 (6.96)              \\ \hline
Moderate/Severe                                                                             & \multicolumn{2}{c|}{40 (1.19)}                & \multicolumn{1}{c|}{27 (1.17)}                & \multicolumn{1}{c|}{16 (1.12)}                & 14 (1.52)              & \multicolumn{1}{c|}{73 (1.25)}               & \multicolumn{1}{c|}{157 (0.82)}                & \multicolumn{1}{c|}{202 (0.66)}                & 163 (0.73)                \\ \hline
\textbf{Diabetes, N (\%)$^\mathrm{All}$}                                                    & \multicolumn{2}{c|}{}                         & \multicolumn{1}{c|}{}                         & \multicolumn{1}{c|}{}                         &                        & \multicolumn{1}{c|}{}                        & \multicolumn{1}{c|}{}                          & \multicolumn{1}{c|}{}                          &                           \\ \hline
Yes, without organ damage                                                                   & \multicolumn{2}{c|}{795 (23.59)}              & \multicolumn{1}{c|}{487 (21.05)}              & \multicolumn{1}{c|}{313 (21.90)}              & 175 (19.04)            & \multicolumn{1}{c|}{1,020 (17.45)}           & \multicolumn{1}{c|}{3,184 (16.63)}             & \multicolumn{1}{c|}{4,640 (15.27)}             & 3,200 (14.28)             \\ \hline
Yes, with organ damage                                                                      & \multicolumn{2}{c|}{202 (5.99)}               & \multicolumn{1}{c|}{121 (5.23)}               & \multicolumn{1}{c|}{75 (5.25)}                & 52 (5.66)              & \multicolumn{1}{c|}{276 (4.72)}              & \multicolumn{1}{c|}{643 (3.36)}                & \multicolumn{1}{c|}{956 (3.15)}                & 591 (2.64)                \\ \hline
\textbf{Hemiplegia / Paraplegia, N (\%)$^\mathrm{All}$}                                     & \multicolumn{2}{c|}{238 (7.06)}               & \multicolumn{1}{c|}{169 (7.30)}               & \multicolumn{1}{c|}{77 (5.39)}                & 60 (6.53)              & \multicolumn{1}{c|}{110 (1.88)}              & \multicolumn{1}{c|}{317 (1.66)}                & \multicolumn{1}{c|}{467 (1.54)}                & 259 (1.16)                \\ \hline
\textbf{Kidney, N (\%)$^\mathrm{All}$}                                                      & \multicolumn{2}{c|}{897 (26.62)}              & \multicolumn{1}{c|}{519 (22.43)}              & \multicolumn{1}{c|}{363 (25.40)}              & 229 (24.92)            & \multicolumn{1}{c|}{963 (16.48)}             & \multicolumn{1}{c|}{2,628 (13.73)}             & \multicolumn{1}{c|}{3,626 (11.93)}             & 2,492 (11.12)             \\ \hline
\textbf{Malignant tumor, N (\%)}                                                            & \multicolumn{2}{c|}{227 (6.74)}               & \multicolumn{1}{c|}{155 (6.70)}               & \multicolumn{1}{c|}{98 (6.86)}                & 57 (6.20)              & \multicolumn{1}{c|}{443 (7.58)}              & \multicolumn{1}{c|}{1,183 (6.18)}              & \multicolumn{1}{c|}{1,953 (6.43)}              & 1,404 (6.27)              \\ \hline
\textbf{Metastatic solid tumor, N (\%)$^{1}$}                                               & \multicolumn{2}{c|}{113 (3.35)}               & \multicolumn{1}{c|}{96 (4.15)}                & \multicolumn{1}{c|}{60 (4.20)}                & 44 (4.79)              & \multicolumn{1}{c|}{273 (4.67)}              & \multicolumn{1}{c|}{734 (3.83)}                & \multicolumn{1}{c|}{1,092 (3.59)}              & 745 (3.33)                \\ \hline
\textbf{HIV, N (\%)$^{2,4}$}                                                                & \multicolumn{2}{c|}{1 (0.03)}                 & \multicolumn{1}{c|}{0 (0.00)}                 & \multicolumn{1}{c|}{1 (0.07)}                 & 0 (0.00)               & \multicolumn{1}{c|}{4 (0.07)}                & \multicolumn{1}{c|}{8 (0.04)}                  & \multicolumn{1}{c|}{28 (0.09)}                 & 22 (0.10)                 \\ \hline
\textbf{Charlson 18, N (\%)$^\mathrm{All}$}                                                 & \multicolumn{2}{c|}{791 (23.47)}              & \multicolumn{1}{c|}{499 (21.56)}              & \multicolumn{1}{c|}{321 (22.46)}              & 194 (21.11)            & \multicolumn{1}{c|}{814 (13.93)}             & \multicolumn{1}{c|}{1,871 (9.77)}              & \multicolumn{1}{c|}{2,579 (8.49)}              & 1,563 (6.98)              \\ \hline
\textbf{Inflammatory bowel disease, N (\%)$^{3,4}$}                                         & \multicolumn{2}{c|}{174 (5.16)}               & \multicolumn{1}{c|}{122 (5.27)}               & \multicolumn{1}{c|}{83 (5.81)}                & 60 (6.53)              & \multicolumn{1}{c|}{315 (5.39)}              & \multicolumn{1}{c|}{799 (4.17)}                & \multicolumn{1}{c|}{1,207 (3.97)}              & 985 (4.40)                \\ \hline
\textbf{Angina, N (\%)$^{2,3,4}$}                                                           & \multicolumn{2}{c|}{250 (7.42)}               & \multicolumn{1}{c|}{167 (7.22)}               & \multicolumn{1}{c|}{97 (6.79)}                & 62 (6.75)              & \multicolumn{1}{c|}{401 (6.86)}              & \multicolumn{1}{c|}{973 (5.08)}                & \multicolumn{1}{c|}{1,432 (4.71)}              & 1,047 (4.67)              \\ \hline
\textbf{Arritmia, N (\%)}                                                                   & \multicolumn{2}{c|}{74 (2.20)}                & \multicolumn{1}{c|}{43 (1.86)}                & \multicolumn{1}{c|}{30 (2.10)}                & 13 (1.41)              & \multicolumn{1}{c|}{89 (1.52)}               & \multicolumn{1}{c|}{264 (1.38)}                & \multicolumn{1}{c|}{378 (1.24)}                & 229 (1.02)                \\ \hline
\textbf{Arterial hypertension, N (\%)$^\mathrm{All}$}                                       & \multicolumn{2}{c|}{2,362 (70.09)}            & \multicolumn{1}{c|}{1,582 (68.37)}            & \multicolumn{1}{c|}{1,011 (70.75)}            & 664 (72.25)            & \multicolumn{1}{c|}{3,350 (57.31)}           & \multicolumn{1}{c|}{10,833 (56.58)}            & \multicolumn{1}{c|}{16,332 (53.75)}            & 11,454 (51.12)            \\ \hline
\textbf{Dyslipidemia, N (\%)$^{4}$}                                                         & \multicolumn{2}{c|}{1,771 (52.55)}            & \multicolumn{1}{c|}{1,169 (50.52)}            & \multicolumn{1}{c|}{695 (48.64)}              & 505 (54.95)            & \multicolumn{1}{c|}{3,040 (52.01)}           & \multicolumn{1}{c|}{9,630 (50.30)}             & \multicolumn{1}{c|}{14,861 (48.91)}            & 11,081 (49.46)            \\ \hline
\textbf{Lymphoma, N (\%)$^{1,2}$}                                                           & \multicolumn{2}{c|}{58 (1.72)}                & \multicolumn{1}{c|}{45 (1.94)}                & \multicolumn{1}{c|}{31 (2.17)}                & 19 (2.07)              & \multicolumn{1}{c|}{218 (3.73)}              & \multicolumn{1}{c|}{596 (3.11)}                & \multicolumn{1}{c|}{1,075 (3.54)}              & 848 (3.78)                \\ \hline
\textbf{Leukemia, N (\%)}                                                                   & \multicolumn{2}{c|}{11 (0.33)}                & \multicolumn{1}{c|}{13 (0.56)}                & \multicolumn{1}{c|}{5 (0.35)}                 & 3 (0.33)               & \multicolumn{1}{c|}{28 (0.48)}               & \multicolumn{1}{c|}{57 (0.30)}                 & \multicolumn{1}{c|}{106 (0.35)}                & 80 (0.36)                 \\ \hline
\textbf{Coagulopathy, N (\%)}                                                               & \multicolumn{2}{c|}{42 (1.25)}                & \multicolumn{1}{c|}{18 (0.78)}                & \multicolumn{1}{c|}{10 (0.70)}                & 11 (1.20)              & \multicolumn{1}{c|}{65 (1.11)}               & \multicolumn{1}{c|}{154 (0.80)}                & \multicolumn{1}{c|}{188 (0.62)}                & 111 (0.50)                \\ \hline
\textbf{Gastrointestinal bleeding, N (\%)$^\mathrm{All}$}                                   & \multicolumn{2}{c|}{152 (4.51)}               & \multicolumn{1}{c|}{92 (3.98)}                & \multicolumn{1}{c|}{55 (3.85)}                & 42 (4.57)              & \multicolumn{1}{c|}{166 (2.84)}              & \multicolumn{1}{c|}{343 (1.79)}                & \multicolumn{1}{c|}{518 (1.70)}                & 326 (1.45)                \\ \hline
\textbf{Asthma, N (\%)$^{1}$}                                                               & \multicolumn{2}{c|}{348 (10.33)}              & \multicolumn{1}{c|}{292 (12.62)}              & \multicolumn{1}{c|}{169 (11.83)}              & 105 (11.43)            & \multicolumn{1}{c|}{823 (14.08)}             & \multicolumn{1}{c|}{2,204 (11.51)}             & \multicolumn{1}{c|}{3,162 (10.41)}             & 2,250 (10.04)             \\ \hline
\textbf{Bronchiectasis, N (\%)}                                                             & \multicolumn{2}{c|}{72 (2.14)}                & \multicolumn{1}{c|}{46 (1.99)}                & \multicolumn{1}{c|}{27 (1.89)}                & 22 (2.39)              & \multicolumn{1}{c|}{149 (2.55)}              & \multicolumn{1}{c|}{352 (1.84)}                & \multicolumn{1}{c|}{492 (1.62)}                & 364 (1.62)                \\ \hline
\textbf{Chronic bronchitis, N (\%)$^{2,3,4}$}                                               & \multicolumn{2}{c|}{389 (11.54)}              & \multicolumn{1}{c|}{275 (11.88)}              & \multicolumn{1}{c|}{181 (12.67)}              & 106 (11.53)            & \multicolumn{1}{c|}{726 (12.42)}             & \multicolumn{1}{c|}{1,745 (9.11)}              & \multicolumn{1}{c|}{2,651 (8.73)}              & 1,751 (7.81)              \\ \hline
\textbf{Cystic fibrosis, N (\%)$^\mathrm{All}$}                                             & \multicolumn{2}{c|}{315 (9.35)}               & \multicolumn{1}{c|}{179 (7.74)}               & \multicolumn{1}{c|}{105 (7.35)}               & 92 (10.01)             & \multicolumn{1}{c|}{190 (3.25)}              & \multicolumn{1}{c|}{409 (2.14)}                & \multicolumn{1}{c|}{547 (1.80)}                & 297 (1.33)                \\ \hline
\textbf{Interstitial lung disease, N (\%)}                                                  & \multicolumn{2}{c|}{17 (0.50)}                & \multicolumn{1}{c|}{10 (0.43)}                & \multicolumn{1}{c|}{4 (0.28)}                 & 5 (0.54)               & \multicolumn{1}{c|}{58 (0.99)}               & \multicolumn{1}{c|}{99 (0.52)}                 & \multicolumn{1}{c|}{157 (0.52)}                & 122 (0.54)                \\ \hline
\textit{Basic treatments}                                                                   & \multicolumn{2}{c|}{}                         & \multicolumn{1}{c|}{}                         & \multicolumn{1}{c|}{}                         &                        & \multicolumn{1}{c|}{}                        & \multicolumn{1}{c|}{}                          & \multicolumn{1}{c|}{}                          &                           \\ \hline
\textbf{Antidiabetics, N (\%)$^{1,2,3}$}                                                    & \multicolumn{2}{c|}{757 (22.46)}              & \multicolumn{1}{c|}{451 (19.49)}              & \multicolumn{1}{c|}{300 (20.99)}              & 166 (18.06)            & \multicolumn{1}{c|}{1,021 (17.47)}           & \multicolumn{1}{c|}{3,188 (16.65)}             & \multicolumn{1}{c|}{4,735 (15.58)}             & 3,339 (14.90)             \\ \hline
\textbf{Cardiovascular, N (\%)$^{3,4}$}                                                     & \multicolumn{2}{c|}{327 (9.70)}               & \multicolumn{1}{c|}{192 (8.30)}               & \multicolumn{1}{c|}{131 (9.17)}               & 91 (9.90)              & \multicolumn{1}{c|}{540 (9.24)}              & \multicolumn{1}{c|}{1,366 (7.14)}              & \multicolumn{1}{c|}{2,021 (6.65)}              & 1,489 (6.65)              \\ \hline
\textbf{Antihypertensive, N (\%)}                                                           & \multicolumn{2}{c|}{78 (2.31)}                & \multicolumn{1}{c|}{58 (2.51)}                & \multicolumn{1}{c|}{39 (2.73)}                & 18 (1.96)              & \multicolumn{1}{c|}{152 (2.60)}              & \multicolumn{1}{c|}{364 (1.90)}                & \multicolumn{1}{c|}{597 (1.96)}                & 371 (1.66)                \\ \hline
\textbf{Diuretics, N (\%)$^\mathrm{All}$}                                                   & \multicolumn{2}{c|}{1,186 (35.19)}            & \multicolumn{1}{c|}{794 (34.31)}              & \multicolumn{1}{c|}{468 (32.75)}              & 292 (31.77)            & \multicolumn{1}{c|}{1,047 (17.91)}           & \multicolumn{1}{c|}{2,860 (14.94)}             & \multicolumn{1}{c|}{3,946 (12.99)}             & 2,430 (10.85)             \\ \hline
\textbf{Beta-blockers, N (\%)$^{4}$}                                                        & \multicolumn{2}{c|}{592 (17.57)}              & \multicolumn{1}{c|}{355 (15.34)}              & \multicolumn{1}{c|}{227 (15.89)}              & 165 (17.95)            & \multicolumn{1}{c|}{939 (16.07)}             & \multicolumn{1}{c|}{2,676 (13.98)}             & \multicolumn{1}{c|}{4,166 (13.71)}             & 3,012 (13.44)             \\ \hline
\textbf{Calcium channel blockers, N (\%)$^\mathrm{All}$}                                    & \multicolumn{2}{c|}{413 (12.26)}              & \multicolumn{1}{c|}{286 (12.36)}              & \multicolumn{1}{c|}{186 (13.02)}              & 107 (11.64)            & \multicolumn{1}{c|}{587 (10.04)}             & \multicolumn{1}{c|}{1,922 (10.04)}             & \multicolumn{1}{c|}{2,785 (9.17)}              & 1,893 (8.45)              \\ \hline
\textbf{System inhibitors, N (\%)$^{1,2}$}                                                  & \multicolumn{2}{c|}{1,178 (34.96)}            & \multicolumn{1}{c|}{775 (33.49)}              & \multicolumn{1}{c|}{498 (34.85)}              & 319 (34.71)            & \multicolumn{1}{c|}{2,303 (39.40)}           & \multicolumn{1}{c|}{7,563 (39.50)}             & \multicolumn{1}{c|}{11,609 (38.21)}            & 8,200 (36.60)             \\ \hline
\textbf{Lipid lowering drugs/statins, N (\%)$^\mathrm{All}$}                                & \multicolumn{2}{c|}{835 (24.78)}              & \multicolumn{1}{c|}{562 (24.29)}              & \multicolumn{1}{c|}{304 (21.27)}              & 247 (26.88)            & \multicolumn{1}{c|}{2,096 (35.86)}           & \multicolumn{1}{c|}{6,777 (35.40)}             & \multicolumn{1}{c|}{10,359 (34.09)}            & 7,806 (34.84)             \\ \hline
\textbf{NSAIDs, N (\%)$^\mathrm{All}$}                                                      & \multicolumn{2}{c|}{203 (6.02)}               & \multicolumn{1}{c|}{111 (4.80)}               & \multicolumn{1}{c|}{99 (6.93)}                & 57 (6.20)              & \multicolumn{1}{c|}{1,103 (18.87)}           & \multicolumn{1}{c|}{3,357 (17.53)}             & \multicolumn{1}{c|}{5,579 (18.36)}             & 4,372 (19.51)             \\ \hline
\textbf{Direct oral anticoagulants, N (\%)$^\mathrm{All}$}                                  & \multicolumn{2}{c|}{1,722 (51.10)}            & \multicolumn{1}{c|}{1,108 (47.88)}            & \multicolumn{1}{c|}{708 (49.55)}              & 453 (49.29)            & \multicolumn{1}{c|}{1,863 (31.87)}           & \multicolumn{1}{c|}{5,563 (29.06)}             & \multicolumn{1}{c|}{8,328 (27.41)}             & 5,448 (24.31)             \\ \hline
\textbf{Antiplatelets, N (\%)$^\mathrm{All}$}                                               & \multicolumn{2}{c|}{950 (28.19)}              & \multicolumn{1}{c|}{658 (28.44)}              & \multicolumn{1}{c|}{369 (25.82)}              & 250 (27.20)            & \multicolumn{1}{c|}{1,067 (18.25)}           & \multicolumn{1}{c|}{2,978 (15.55)}             & \multicolumn{1}{c|}{4,539 (14.94)}             & 3,289 (14.68)             \\ \hline
\textbf{Heparin, N (\%)$^\mathrm{All}$}                                                     & \multicolumn{2}{c|}{151 (4.48)}               & \multicolumn{1}{c|}{119 (5.14)}               & \multicolumn{1}{c|}{111 (7.77)}               & 38 (4.13)              & \multicolumn{1}{c|}{109 (1.86)}              & \multicolumn{1}{c|}{595 (3.11)}                & \multicolumn{1}{c|}{984 (3.24)}                & 290 (1.29)                \\ \hline
\textbf{Azitromizine, N (\%)$^{1,3}$}                                                       & \multicolumn{2}{c|}{191 (5.67)}               & \multicolumn{1}{c|}{36 (1.56)}                & \multicolumn{1}{c|}{27 (1.89)}                & 6 (0.65)               & \multicolumn{1}{c|}{120 (2.05)}              & \multicolumn{1}{c|}{225 (1.18)}                & \multicolumn{1}{c|}{321 (1.06)}                & 166 (0.74)                \\ \hline
\textbf{Broncodilators, N(\%)$^{1}$}                                                        & \multicolumn{2}{c|}{351 (10.42)}              & \multicolumn{1}{c|}{287 (12.40)}              & \multicolumn{1}{c|}{172 (12.04)}              & 99 (10.77)             & \multicolumn{1}{c|}{865 (14.80)}             & \multicolumn{1}{c|}{2,344 (12.24)}             & \multicolumn{1}{c|}{3,450 (11.36)}             & 2,418 (10.79)             \\ \hline
\textbf{Immunosuppressants, N (\%)$^{1,2,4}$}                                               & \multicolumn{2}{c|}{21 (0.62)}                & \multicolumn{1}{c|}{15 (0.65)}                & \multicolumn{1}{c|}{11 (0.77)}                & 4 (0.44)               & \multicolumn{1}{c|}{123 (2.10)}              & \multicolumn{1}{c|}{343 (1.79)}                & \multicolumn{1}{c|}{509 (1.68)}                & 487 (2.17)                \\ \hline
\textbf{Chronic systemic steroids, N (\%)$^{1,2}$}                                          & \multicolumn{2}{c|}{340 (10.09)}              & \multicolumn{1}{c|}{122 (5.27)}               & \multicolumn{1}{c|}{73 (5.11)}                & 46 (5.01)              & \multicolumn{1}{c|}{348 (5.95)}              & \multicolumn{1}{c|}{763 (3.99)}                & \multicolumn{1}{c|}{1,226 (4.04)}              & 901 (4.02)                \\ \hline
\textit{Output variables}                                                                   & \multicolumn{2}{c|}{}                         & \multicolumn{1}{c|}{}                         & \multicolumn{1}{c|}{}                         &                        & \multicolumn{1}{c|}{}                        & \multicolumn{1}{c|}{}                          & \multicolumn{1}{c|}{}                          &                           \\ \hline
\textbf{Hospitalization, N (\%)$^{1,4}$}                                                    & \multicolumn{2}{c|}{605 (17.95)}              & \multicolumn{1}{c|}{494 (21.35)}              & \multicolumn{1}{c|}{289 (20.22)}              & 95 (10.34)             & \multicolumn{1}{c|}{3,229 (55.24)}           & \multicolumn{1}{c|}{4,244 (22.17)}             & \multicolumn{1}{c|}{6,217 (20.46)}             & 1,296 (5.78)              \\ \hline
\textbf{Death, N (\%)$^\mathrm{All}$}                                                       & \multicolumn{2}{c|}{827 (24.54)}              & \multicolumn{1}{c|}{556 (24.03)}              & \multicolumn{1}{c|}{285 (19.94)}              & 98 (10.66)             & \multicolumn{1}{c|}{800 (13.69)}             & \multicolumn{1}{c|}{1,134 (5.92)}              & \multicolumn{1}{c|}{1,511 (4.97)}              & 437 (1.95)                \\ \hline
\textbf{Bad progress, N (\%)$^\mathrm{All}$}                                                & \multicolumn{2}{c|}{835 (24.78)}              & \multicolumn{1}{c|}{557 (24.07)}              & \multicolumn{1}{c|}{285 (19.94)}              & 100 (10.88)            & \multicolumn{1}{c|}{1,017 (17.40)}           & \multicolumn{1}{c|}{1,452 (7.58)}              & \multicolumn{1}{c|}{2,138 (7.04)}              & 524 (2.34)                \\ \hline
\end{tabular}
\end{scriptsize}
\end{table}
\end{landscape}
\clearpage
\end{appendices}

\end{document}